\documentclass{emulateapj}
\renewcommand\email\texttt
\shorttitle{The Luminosity Function of the Milky Way Satellites}
\shortauthors{Koposov et al.}
\def\Ss{S_{\rm star}}
\def\Sg{S_{\rm gal}}
\def\rh{r_{\rm h}}
\def\change#1{{#1}}
\begin{document}
\title{The Luminosity Function of the Milky Way Satellites}
\author{S. Koposov\altaffilmark{1,2},
V. Belokurov\altaffilmark{2},
N.W. Evans\altaffilmark{2}, 
P.C. Hewett\altaffilmark{2},
M.J. Irwin\altaffilmark{2},
G. Gilmore\altaffilmark{2},
D.B. Zucker\altaffilmark{2},
H.-W. Rix\altaffilmark{1},
M. Fellhauer\altaffilmark{2}, 
E.F. Bell\altaffilmark{1},
E.V. Glushkova\altaffilmark{3}
}
\altaffiltext{1}{Max Planck Institute for Astronomy, K\"{o}nigstuhl
17, 69117 Heidelberg, Germany}
\altaffiltext{2}{Institute of Astronomy, Madingley Road, Cambridge CB3 0HA,
UK;\email{koposov,vasily,nwe@ast.cam.ac.uk}}
\altaffiltext{3}{Sternberg Astronomical Institute, Universitetskiy pr.,
13, 119992, Moscow, Russia}
\begin{abstract}
  We quantify the detectability of stellar Milky Way
  satellites in the Sloan Digital Sky Survey (SDSS) Data Release 5.
  We show that the effective search volumes for the recently discovered 
  SDSS--satellites depend strongly on their luminosity, with their
  maximum distance, $D_{max}$, substantially
  smaller than the Milky Way halo's virial radius.
  Calculating the maximum accessible volume, $V_{max}$, for all faint
  detected satellites, allows the calculation of
  the luminosity function for Milky Way satellite galaxies,
  accounting quantitatively for their detectability.
  We find that the number density of satellite galaxies continues to rise
  towards low luminosities, but may flatten at $M_V \sim -5$; within
  the uncertainties, the luminosity function can be described by a
  single power law $dN/dM_{V}= 10 \times 10^{0.1 (M_V+5)}$, spanning
  luminosities from $M_V=-2$ all the way to the luminosity of the Large
  Magellanic Cloud.
  Comparing these results to several semi-analytic galaxy
  formation models, we find that their predictions differ significantly from
  the data: either the shape of the luminosity function, or the surface
  brightness distributions of the models, do not match.
\end{abstract}
\keywords{Galaxy: halo -- Galaxy: structure -- Galaxy: formation -- Local Group}
\section{Introduction}
\begin{figure}
 \plotone{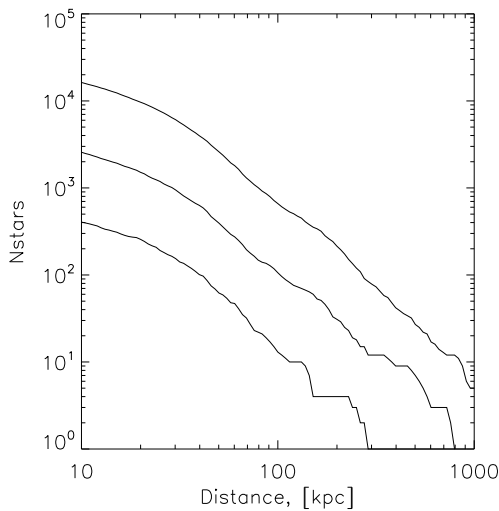}
 \caption{The number of stars brighter than $r \simeq22.5 $ in random
   realizations of Milky Way satellites of luminosity $M_r \sim -3, -5,    
   -7$ (from bottom to top) with M92-like stellar populations, as a
   function of distance from the Sun. The approximate number of stars
   required for a significant detection (by the algorithm described in
   Section~\ref{algorithm_section}) is $\simeq$30.}
 \label{nstars_vs_dist}
\end{figure}

In Cold Dark Matter (CDM) models, large spiral galaxies like the Milky
Way and M31 form within extensive dark matter halos from the merging
and accretion of smaller systems.  Although CDM models have had many
successes on larger scales, one of the most serious challenges facing
CDM models is the so-called ``missing satellite'' problem. First
identified by~\citet{Kl99} and \citet{Mo99}, the problem manifests
itself through the prediction by CDM models of at least 1-2 orders of
magnitude more low-mass sub-halos at the present epoch compared to the
observed abundance of dwarf galaxies surrounding the Milky Way and
M31.

There have been a number of theoretical proposals to solve this
problem. For example, the satellites that are observed could be
embedded only in the rarer, more massive dark sub-halos~\citep{Sto02},
or, the satellites may form only in the rare peaks of halos that were
above a given mass at reionization~\citep{Di05,Mo06}. Alternatively,
star formation in low mass systems could be inhibited by
photoionization in the early Universe~\citep{Bu01,So02,Be02}. All
these ideas do not alter the abundance of dark matter sub-halos, but
propose to solve the observed discrepancy by producing a smaller
number of directly observable satellites, thus breaking any simple
relationship between mass and luminosity.

The known Milky Way dwarf spheroidal (dSph) satellites have been
discovered by a variety of methods. The first seven were discovered
serendipitously by visual inspection of photographic plates, the
Sextans dSph was found using automated scans of photographic plates
and the Sagittarius dSph in a radial velocity survey of the Milky Way
bulge.  All-sky photographic surveys cover most of the sky away from
the Zone of Avoidance, but searches of plates are limited to surface
brightnesses of $\sim 25.5$ mag arcsec$^{-2}$ ~\citep{Wh07}.  The
sample of known dSphs has long been bedeviled with selection effects,
which are difficult to model with any accuracy. This situation has
changed recently with the advent of very large area, homogeneous,
photometric surveys such as the Sloan Digital Sky Survey (SDSS; York
et al. 2000). The SDSS makes it possible to carry out a systematic
survey for satellite galaxies, which are detectable through their
resolved stellar populations down to extremely low surface
brightnesses. In essence, SDSS greatly \change{facilitates} systematic
searches for overdensities of stars in position-color-magnitude space.
\begin{figure}
\begin{center}
 \includegraphics[height=6cm]{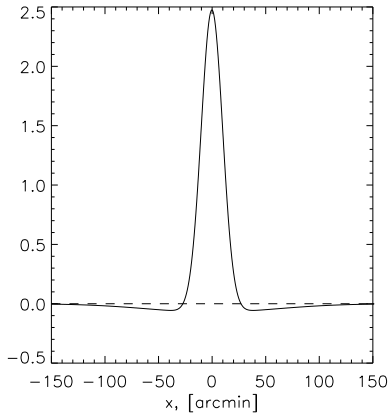}
\end{center}
 \caption{Differential convolution kernel applied to the stellar
   catalog to identify overdensities of a particular scale. A
   one-dimensional slice of the two-dimensional kernel is shown, where
   the width, or $\sigma$, of the inner Gaussian is 6\arcmin\ and of
   the outer Gaussian is 60\arcmin.}
 \label{kernel}
\end{figure}

\citet{Wi02} carried out the first SDSS--based survey for resolved
Milky Way satellites, subsequently discovering a new dwarf galaxy,
Ursa Major \citep{Wi05a} as well as an unusually large globular
cluster, Willman 1~\citep{Wi05b} \change{ -- although later evidence
may favor its interpretation as a dark matter dominated dwarf galaxy
with multiple stellar populations \citep{Martin07}}.  The color image
``Field of Streams''~\citep{Be06}, composed of magnitude slices of the
stellar density in the SDSS around the North Galactic Cap, proved to
be a treasure-trove for dwarf galaxies, as Canes Venatici, Bootes~I
and Ursa Major~II~\citep{Zu06a,Zu06b,Be06b} were all found in quick
succession. A systematic search in the ``Field of Streams'' led to the
discovery of five more satellite galaxies, Canes Venatici~II, Leo~IV,
Hercules, Coma, and Leo T, as well as another large globular cluster,
Segue 1~\citep{Be07,Ir07}. Very recently, \citet{Wa07} discovered
another low luminosity satellite, Bootes~II.

As the faintest Milky Way satellites currently constitute our best
markers of sub-halos, the faint end of the satellite luminosity
function of the Milky Way satellites can provide stringent constraints
on the process of galaxy formation, and can distinguish between a
number of dark matter, structure formation and reionization
models. So, it is important not merely to carry out a systematic
survey of the star overdensities in SDSS data for the discoveries per
se, but also to compute the detection limits. These detection limits
are the basis for a volume corrected luminosity function estimate and
ultimately for a quantitative connection of satellite frequency and
sub-halo abundance. Such is the purpose of this paper.  It is
important to note that for a volume-corrected estimate of the
luminosity function, it is not necessary to use exactly the same
detection algorithms as \citet{Be07} or \citet{Wi05a}. Similarly, the
detection scheme does not need to be optimal for every individual
dwarf galaxy.  Provided the automated algorithm is able to detect all
the Milky Way satellites, and the completeness properties of the
algorithm are quantified, an estimate of the true luminosity function
can be derived.

\section{Detection of Satellite Galaxy Candidates in SDSS DR5}
\label{algorithm_section}
The SDSS Data Release 5 (DR5) covers $\sim 1/5$ of the sky, or $\sim
8000$ square degrees around the North Galactic Pole. SDSS imaging data
are produced in five photometric bands, $u$, $g$, $r$, $i$, and
$z$~\citep{Fu96,Gu98,Ho01,Am06,Gu06}.  The data are automatically
processed through pipelines to measure photometric and astrometric
properties \citep{Lu99,St02,Pi03,Tu06}.  All magnitudes quoted in this
paper have been corrected for reddening due to Galactic extinction
using the maps of~\citet*{Sc98}. Sometimes it is convenient to report
our results in the $V$ band, for which we use the transformation
$V=g-0.55(g-r)-0.03$ given by~\citet{Sm02}.

The SDSS data with the source catalogs used in this paper was
downloaded from the SAI CAS Virtual Observatory data
center\footnote{\url{http://vo.astronet.ru}}\citep{koposov_bartunov}
and was stored locally in the PostgreSQL database. To perform
queries rapidly on the large dataset, we used the Q3C plugin for the
spatial queries \citep{q3c}.

All the recent SDSS discoveries of dSph around the Milky Way, bar Leo
T, are not directly visible in the flux-limited images, but were
detected as overdensities of resolved stars within certain magnitude
and color ranges. This makes it straightforward to automate a
detection method and assess its efficiency.  The essence of any
detection algorithm is to count the number of stars in a certain
(angular) region on the sky, satisfying specified color and magnitude
criteria, and compare the number to the background value. The excess
of stars depends on the satellite's luminosity and distance.  For a
given luminosity, the distance fixes the number of stars brighter than
the SDSS limiting magnitude, which is given by an integral over the
stellar luminosity function. A simple illustration of the
detectability of objects with a luminosity function like that of M92
is shown in Figure~\ref{nstars_vs_dist}. The curves show the number of
stars brighter than $r = 22.5$ for satellites of three different
absolute magnitudes. The maximal distance probed by surveys like SDSS
is controlled by the apparent magnitude of the brightest stars in the
satellite.  For intrinsically luminous objects, like CVn~I ($M_V =
-7.9$) , we can detect stars at the tip of the red giant branch at
distances of up to $\sim$1\,Mpc. However, for satellites with many
fewer stars, like Hercules ($M_V = -5.7$), the giant branch tip is
simply not populated and we can only detect objects at distances up to
$\sim$300\,kpc.

To identify the excess number of stars associated with a satellite, 
a common approach is to
convolve the spatial distribution of the data with window functions or
filters~\footnote{This idea has a long history, particularly in
  algorithms for searching for features and clusters in imaging data.
  Widely used in astronomy are kernel-based density estimation
  methods, in which the density is obtained by convolving all the data
  points (interpreted as delta-functions) with smoothly decaying
  kernels, which can be Gaussians \citep[see e.g.][]{silverman}. A
  variant of this is used for feature detection in digital images in
  so-called scale-space science
  \citep{lindenberg93,lindenberg98,babaud}.}. To estimate the star
density on different scales, we use a Gaussian of width $\sigma$, that
is,
\begin{equation}
L(x,y,\sigma)=I(x,y)*g(x,y,\sigma), 
\end{equation}
where
\begin{equation}
g(x,y,\sigma) = \frac{1}{2\pi
  \sigma^2}\exp\left(-\frac{x^2+y^2}{2\sigma^2}\right)
\end{equation}
and $I(x,y)$ is the distribution of sources
\begin{equation}
 I(x,y) = \sum\limits_i \delta(x-x_i,y-y_i)
\end{equation}

This allows us to see the stellar density distribution at different
spatial scales. For example, structures with a characteristic size of
1\arcmin\ will be more prominent when the stellar map is convolved
with a 1\arcmin\ kernel, and less prominent when the map is convolved
with 10\arcmin\ and 0.1\arcmin\ kernels. The resulting ``blobs'', or
overdensities, can be easily identified on the differential image
maps, namely
\begin{eqnarray}
\Delta L  & = & L(x, y, \sigma_1) - L (x, y, \sigma_2) \nonumber \\
          & = & I(x,y) * (g(x,y,\sigma_1)-g(x,y,\sigma_2))
\label{eq:convolve}
\end{eqnarray}
Such differential image maps are generally convolutions of the
original distribution with the kernel, which is a difference of two
Gaussians.  A one-dimensional slice of a kernel is shown in
Figure~\ref{kernel}.  When we convolve the map $I(x,y)$ with such a
kernel, we obtain an estimate of the local density minus an estimate of
the local background ($L(x,y,\sigma_2)$). This interpretation allows us to
quantify the
significance as
\begin{eqnarray}
S(x,y,\sigma_1,\sigma_2) & = & {\frac{\Delta L}{\sigma_L}}
\end{eqnarray}
where $\sigma_L^2$ is the variance of $L(x, y, \sigma_1)$.
\begin{eqnarray}
\sigma_L^2 & = & Variance(L(x, y, \sigma_1))=\nonumber \\
&=& Variance(I(x, y)*g(x, y,
\sigma_1))= I(x,y)*g^2(x, y , \sigma_1) =\nonumber \\
 &= & \sum\limits_{i,j} I(x_i,y_j)\,g^2(x-x_i,y-y_j,\sigma_1)\approx \nonumber \\
& \approx & \sum\limits_{i,j} L(x,y,\sigma_2)\,g^2(x- x_i,y-y_j,\sigma_1)=
\nonumber\\
 &= & L(x, y, \sigma_2)\int\int g^2(x, y, \sigma_1)\,dxdy = \frac{L(x, y, \sigma_2)}{4 \pi \sigma_1^2}
\end{eqnarray}
\begin{eqnarray}
S(x,y,\sigma_1,\sigma_2) & = \sqrt{4 \pi}
\sigma_1 \frac{\Delta L} {\sqrt{L(x,y,\sigma_2)}},
\label{eq:sigmaL}
\end{eqnarray}
Under the
assumption that $\sigma_2 >> \sigma_1$ and a Poisson distribution of
the initial set of datapoints, the variance of $S(x,y)$ is
unity. This fact allows us to use the map of $S(x,y)$ to identify 
overdensities above a specified significance threshold.

\begin{figure}
\begin{center}
 \includegraphics[height=4cm]{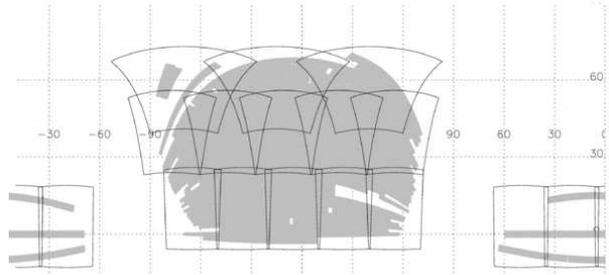}
\end{center}
 \caption{The segmentation of the DR5 area into 17 $32^\circ \times
   32^\circ$ fields, used for the stellar overdensity search described in
   the text.}
 \label{segmentation}
\end{figure}
\begin{figure*}
\begin{center}
 \includegraphics[height=8cm]{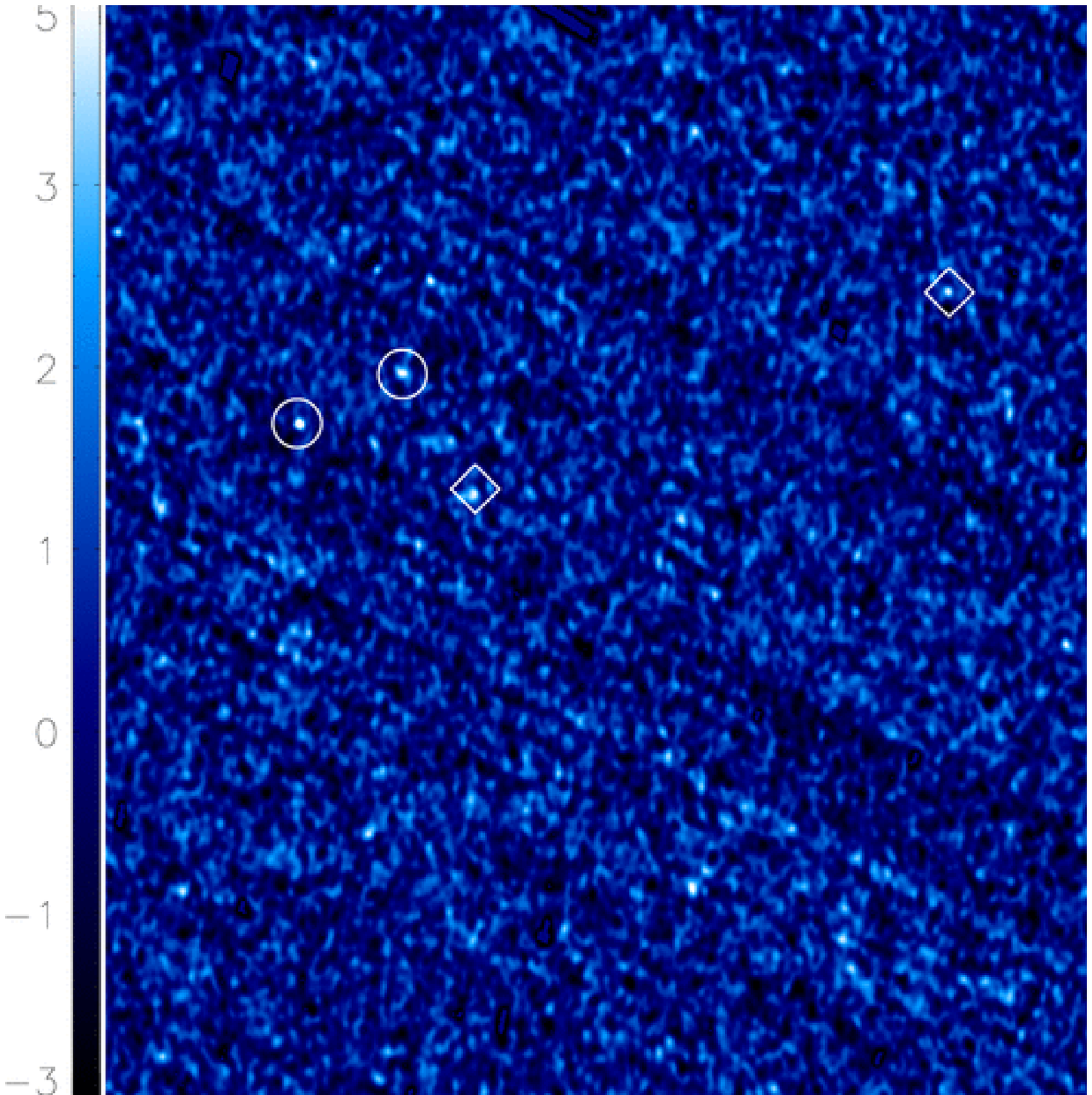}
 \includegraphics[height=8cm]{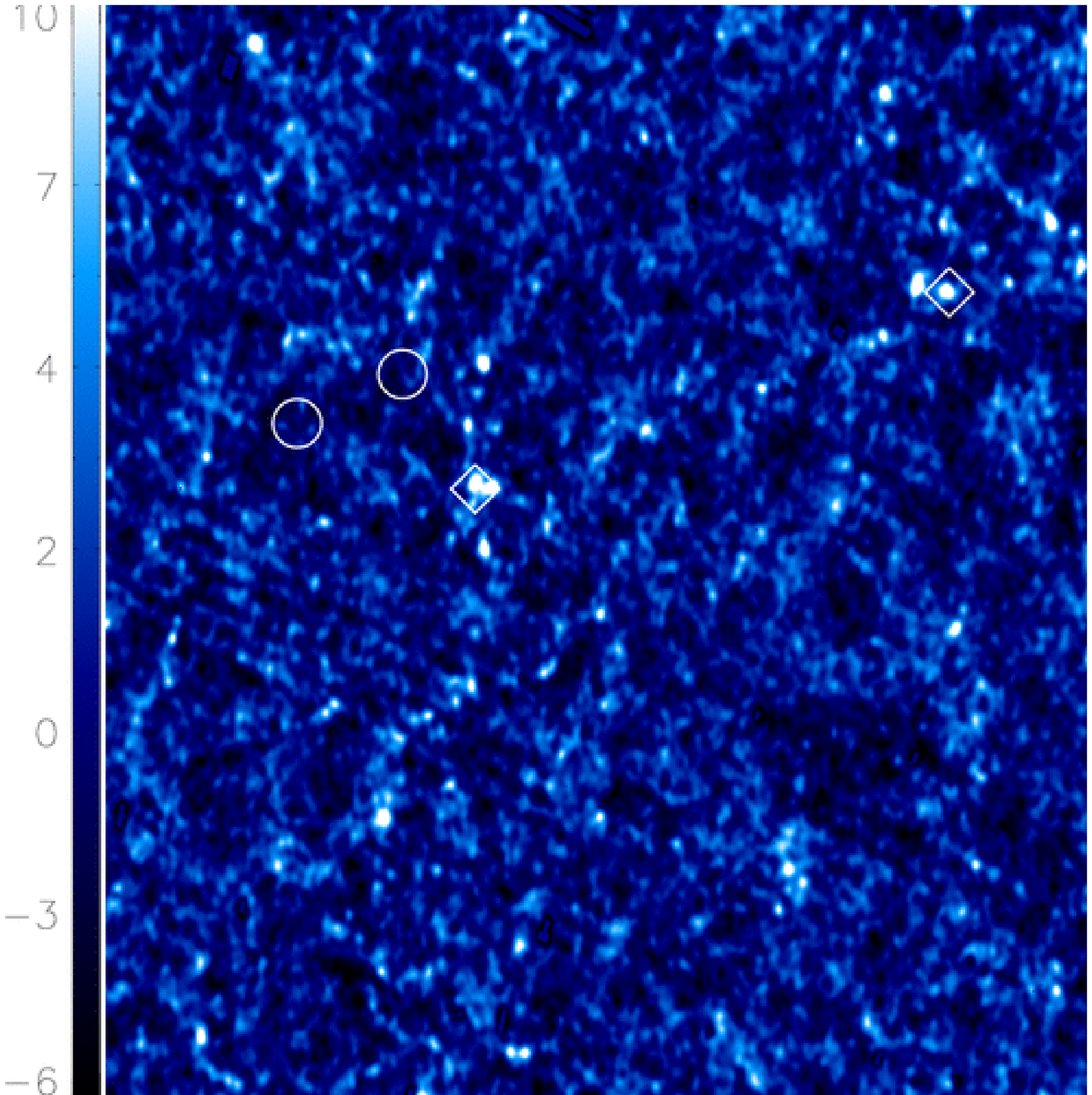}
\end{center}
 \caption{A sample $22^\circ \times 22^\circ$ area in the convolved maps of
   the SDSS DR5 stellar (left) and galaxy (right) catalogs. The
   positions of objects Ursa Major I and Willman 1 are marked by
   circles. The positions of galaxy clusters Abell 773 and Abell 1000
   are marked by diamonds, and demonstrate that galaxy clusters may
   lead to significant peaks in the stellar map. The linear diagonal structures
   seen in both images are caused by SDSS stripes. The images were produced
   using a kernel specified by $\sigma_1 = 4'$ and $\sigma_2=60'$. We reject
   peaks in the convolved stellar map if they coincide with significant peaks 
   in the galaxy distribution.}
 \label{star_gal_maps}
\end{figure*}
\begin{figure}
 \includegraphics[height=4.2cm]{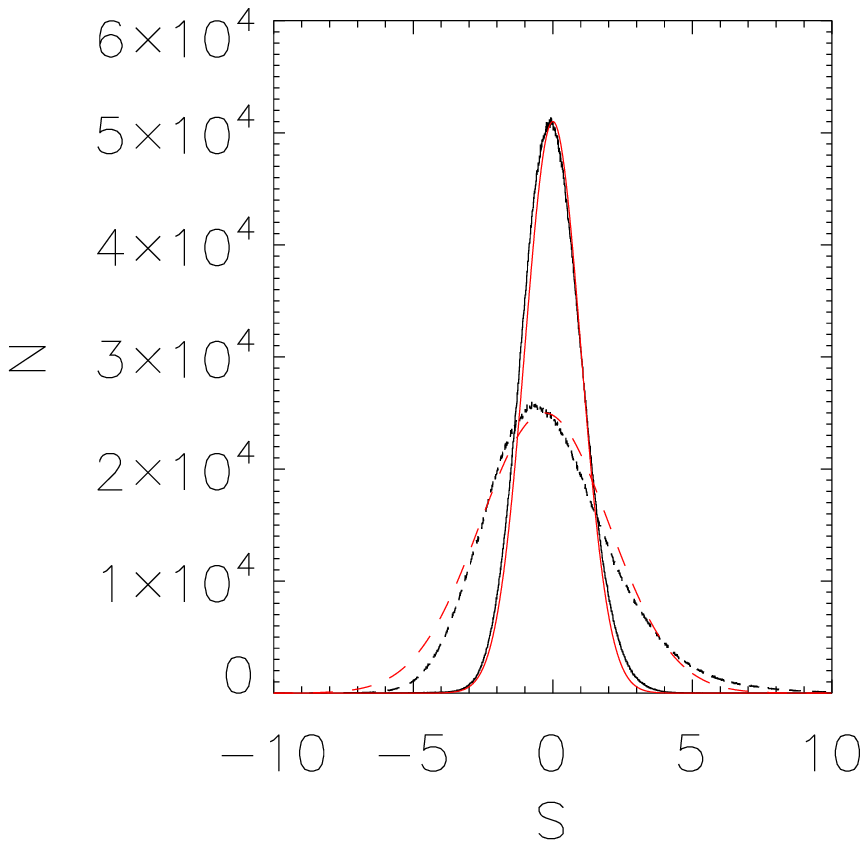}
 \includegraphics[height=4.2cm]{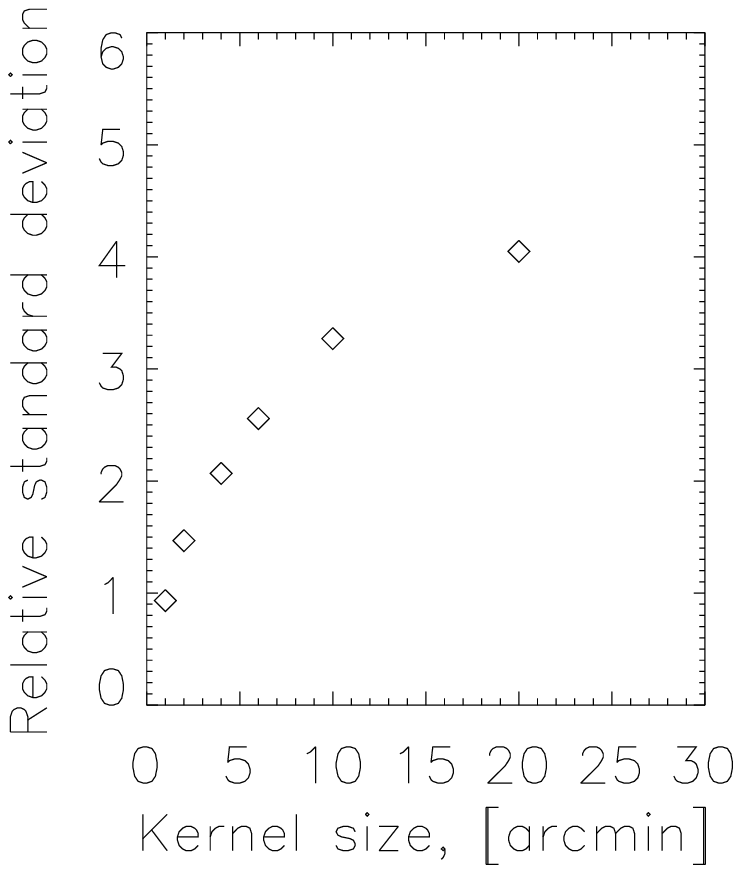}
 \caption{Left: The distribution of pixel values in the convolved star map
   (solid line) and galaxy map (dashed line) for one of our 17 fields in DR5.
   The Gaussian model curves
   with width of 1.0 and 2.3 centered on zero are shown in red. Right:
   The standard deviation in the galaxy map normalized by a
   Poissonian standard deviation as a function of kernel size.}
 \label{pix_distrib}
\end{figure}
\begin{deluxetable}{ccccc}
\tablecaption{Objects Detected and Their Significances \label{tab:zzz}}
\tablehead{ \colhead{Right ascension} & 
\colhead{Declination} & \colhead{$\Ss$} & \colhead{$\Sg$} & 
\colhead{Name} }
\startdata 
205.539 & 28.382 & 170.13 & 8.24 & NGC 5272\\
168.355 & 22.148 & 170.08 & 19.16 & Leo II\\
198.220 & 18.159 & 165.00 & 16.48 & NGC5024\\
152.100 & 12.289 & 123.86 & 14.94 & Leo I\\
199.104 & 17.696 & 122.09 & 5.59 & NGC 5053\\
211.359 & 28.527 & 121.91 & 5.61 & NGC 5466\\
229.006 & -0.130 & 115.10 & 19.48 & Pal5\\
260.038 & 57.914 & 100.02 & 15.64 & Draco\\
250.426 & 36.467 & 94.10 & 2.80 & NGC 6205\\
322.483 & 12.147 & 87.58 & 13.30 & NGC7078\\
182.516 & 18.544 & 79.40 & 7.58 & NGC 4147\\
260.008 & 57.765 & 75.07 & 11.27 & Draco\\
114.534 & 38.873 & 72.28 & 6.80 & NGC 2419\\
323.212 & -0.865 & 65.52 & -1.75 & NGC 7089\\
187.670 & 12.395 & 59.69 & 3.25 & NGC 4486\\
202.011 & 33.549 & 44.12 & 1.68 & CVn I\\
187.419 & 8.003 & 37.18 & 0.93 & NGC 4472\\
149.834 & 30.742 & 28.13 & 12.65 & Leo A\\
190.698 & 2.682 & 27.73 & 5.51 & NGC 4636\\
114.608 & 21.581 & 25.88 & -6.25 & NGC 2420\\
259.027 & 43.063 & 22.76 & -5.60 & NGC 6341\\
183.904 & 36.310 & 20.43 & 7.49 & NGC 4214\\
185.036 & 29.286 & 17.28 & 0.63 & NGC 4278\\
210.010 & 14.503 & 16.95 & 0.32 & Boo I\\
190.773 & 11.598 & 16.84 & -1.52 & NGC 4647/4637/4638\\
186.368 & 12.909 & 14.28 & 0.76 & NGC 4374\\
178.814 & 23.371 & 13.90 & 19.34 & Abell 1413\\
186.444 & 33.539 & 13.10 & 16.06 & NGC 4395\\
148.904 & 69.081 & 13.08 & 19.85 & NGC 3031\\
162.325 & 51.051 & 13.07 & 0.11 & Willman 1\\
242.741 & 14.956 & 12.52 & -0.07 & Pal 14\\
186.315 & 18.181 & 11.59 & -3.72 & NGC 4382\\
132.830 & 63.124 & 11.40 & 0.92 & UMa II\\
186.745 & 23.913 & 11.22 & -1.02 & Coma Berenices\\
143.721 & 17.058 & 10.96 & 3.85 & Leo T\\
188.911 & 12.544 & 10.91 & -0.97 & NGC 4552\\
210.691 & 54.332 & 10.67 & 13.53 & NGC 5457\\
151.369 & 0.070 & 10.64 & 4.11 & Pal 3\\
186.109 & 7.294 &  9.85 & -0.76 & NGC 4365\\
172.319 & 28.961 & 9.53 & 0.26 & Pal 4\\
247.764 & 12.789 & 8.91 & 1.05 & Hercules\\
197.870 & -1.335 & 8.23 & 22.00 & Abell 1689\\
194.292 & 34.298 & 7.39 & -4.61 & CVn II\\
196.743 & 46.569 & 7.30 & 8.29 & Abell 1682\\
193.379 & 46.415 & 6.71 & -2.41 & Candidate X\\
168.146 & 43.440 & 6.52 & -3.22 & Candidate Y\\
352.182 & 14.714 & 6.39 & 1.47 & Pegasus\\
202.388 & 58.404 & 6.37 & -0.59 & NGC 5204\\
225.323 & 1.672 & 6.34 & 1.36 & NGC 5813\\
187.038 & 44.090 & 6.13 & -2.49 & NGC 4449\\
173.235 & -0.554 & 6.10 & 2.92 & Leo IV\\
0.807 & 16.097 & 6.09 & -2.66 & NGC 7814\\
179.144 & 21.049 & 6.05 & 1.14 & Candidate Z\\
184.843 & 5.786 & 6.04 & -1.25 & NGC 4261\\
149.993 & 5.316 & 6.03 & 2.12 & Sextans B\\
139.470 & 51.718 & 5.97 & 17.86 & Abell 773\\
179.223 & 23.379 & 5.96 & 5.22 & galaxy cluster\\
158.695 & 51.918 & 5.95 & 3.72 & UMa I
\enddata
\end{deluxetable}
\begin{figure}
 \includegraphics[height=8cm]{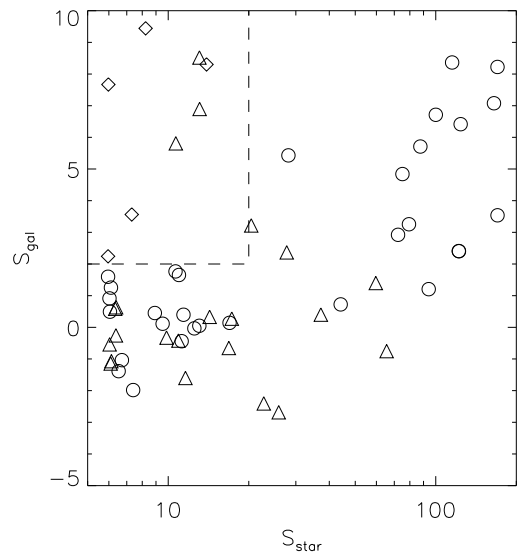}
 \caption{Distribution of Milky Way satellite detections in the 
   $\Ss$ versus $\Sg$ plane. The
   circles mark the known Milky Way satellites, the triangles are RC3
   galaxies, and the rhomboids are galaxy clusters. Objects towards the top
   left of the figure are likely the result of contamination by galaxy 
   clusters or spatially extended galaxies. The decision
   boundary is shown as a dashed line; objects to the right and below the
   dashed line are selected as candidate satellites.}
 \label{sgalversussstar}
\end{figure}
\begin{figure*}
 \includegraphics[height=6cm]{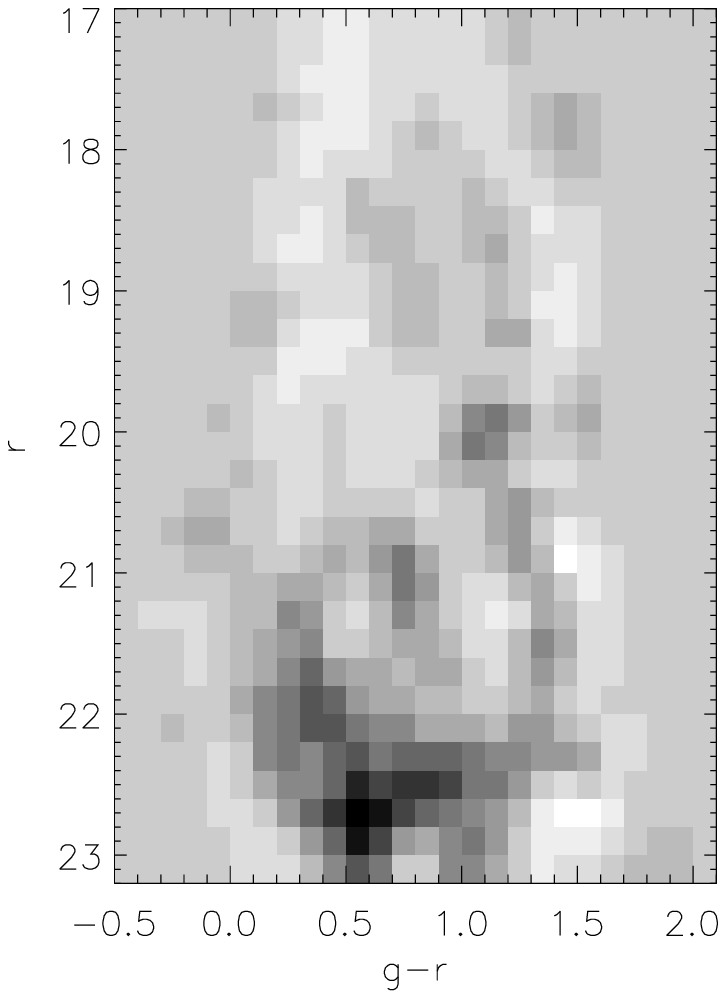}
 \includegraphics[height=6cm]{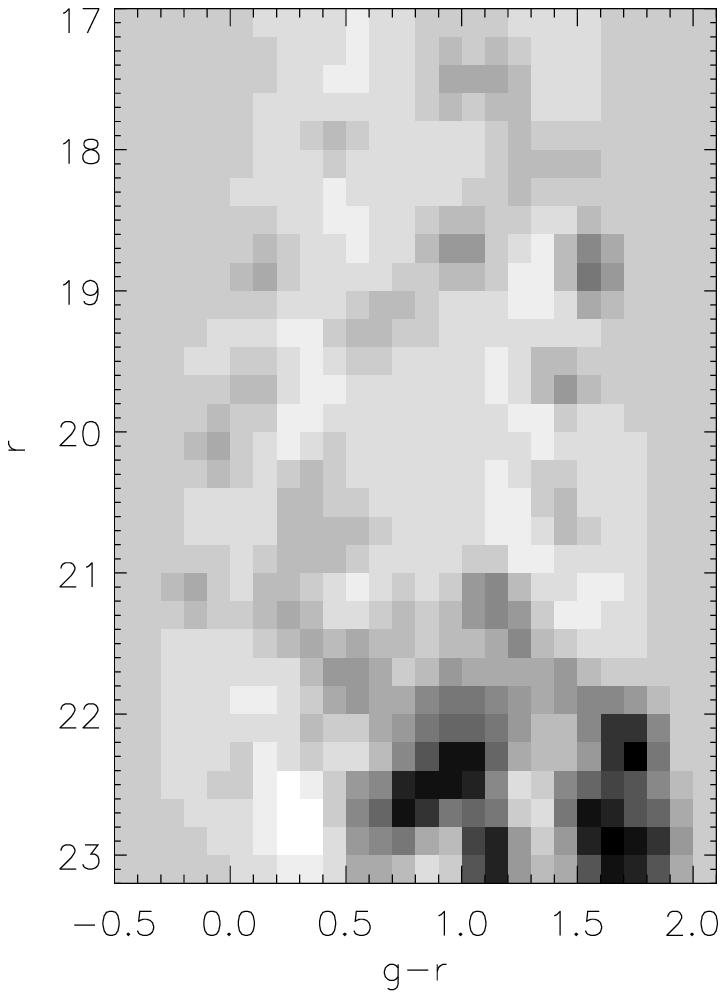}
 \includegraphics[height=6cm]{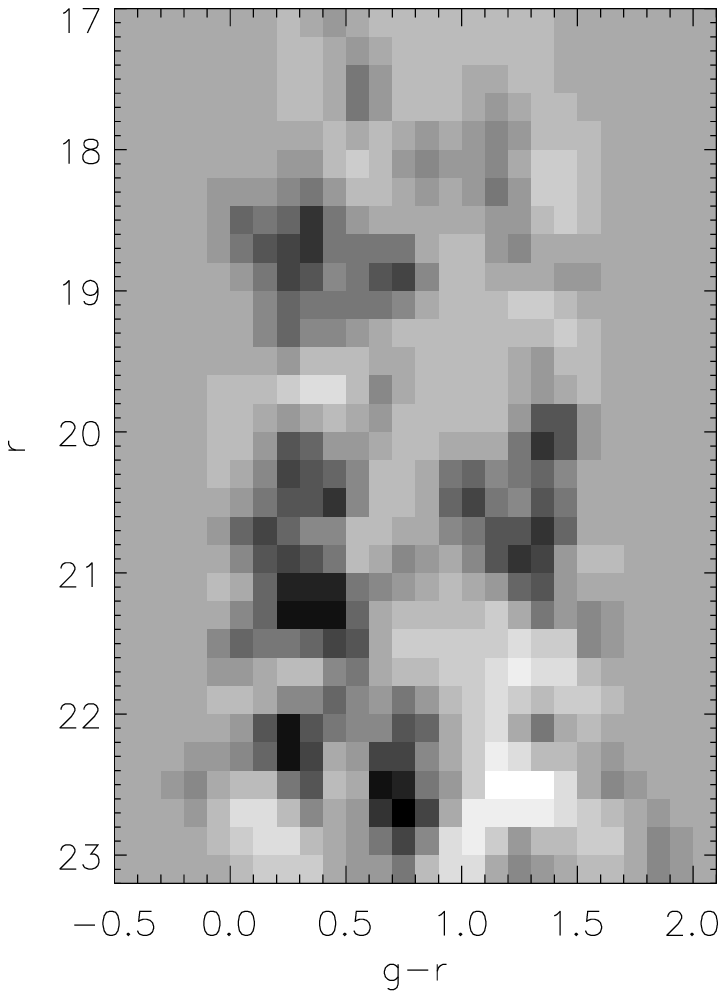}
 \caption{Left to right: Hess diagrams for Candidates X, Y and Z listed in
   Table~\ref{tab:zzz}}
 \label{hess}
\end{figure*}

\section{Application to SDSS data}
\label{observ_section}
SDSS's morphological parameters \citep{Lu01} derived from the imaging
data allow robust discrimination between stars and galaxies down to $r
= 21.5$.  For $21.5 < r \lesssim 22.5$, the discrimination is still
reasonably reliable, but it becomes increasingly untrustworthy below
$r = 22.5$.  Moreover, the catalog is 95$\%$ complete at $r =
22.2$~\citep{St02} and drops quickly below this magnitude.  At the
faint end, the ``stellar'' catalog of unresolved sources is polluted
by faint galaxies which are intrinsically strongly clustered. We will
see shortly that the main task in providing a clean sample of dwarf
galaxy candidates is removal of the extragalactic contaminants, for
which we will employ the SDSS galaxy catalog.

To proceed with the convolution (Eq.~\ref{eq:convolve}), the DR5 field
of view is split into 17 segments as shown in
Figure~\ref{segmentation}.  The division is for computational
convenience and to minimize distortion in the gnomonic projections.
In practice, we select stars and galaxies with a magnitude cut-off of
$r < 22.5$. Due to the properties of the kernel, we expect edge
effects at the boundaries of the DR5 footprint and we discard all
overdensities within 1$^\circ$ of a boundary. We use a color-cut of
$g-r < 1.2$ and kernel sizes with $\sigma_1 = 4'$ and $\sigma_2
=60'$. The color cut is chosen to be as conservative as possible as
regards inclusion of the tip of the red giant branch stars for
metal-poor populations, whilst the kernel size is of the order of the
angular size of the known dwarfs (see next section for details). The
color magnitude cut used in this work may not be optimal for the
detection of each individual dwarf galaxy (e.g. the isochrone masks
should definitely work better), but the primary goal here is not to
define an optimal algorithm, but rather to develop a consistent
algorithm that can detect known objects, for which the detection
efficiency can be determined.  Figure~\ref{star_gal_maps} provides an
example of the application of the detection pipeline to the stellar
and galaxy catalogs of SDSS DR5. The method successfully removes the
varying background to leave underdensities (black regions) and
overdensities (white regions). The sample field of view chosen for
Figure~\ref{star_gal_maps} contains the already known Milky Way
satellites Willman 1 and Ursa Major I \citep{Wi05a,Wi05b}. They are
both recovered in the stellar map with significances of $\Ss = 13.07$
and $5.95$ respectively. However, as we see in
Figure~\ref{star_gal_maps}, unresolved sources in rich galaxy clusters
such as Abell 773 and 1000, visible as prominent overdensities in the
galaxy map, also show up in the stellar map as significant peaks.

In order to remove false positives caused by galaxy clustering,
we need to understand the significance $\Sg$ of
overdensities in the map derived from the galaxy
catalog. Equation~(\ref{eq:sigmaL}) does not hold, because the
underlying distribution is no longer Poissonian
(Figure~\ref{pix_distrib}).  The left panel shows the distributions of
$\Ss$ and $\Sg$ for all pixels in the same field of view as
Figure~\ref{star_gal_maps}. For the stars, the convolved source count
distribution is almost a Gaussian with unit standard deviation, whilst
the distribution for the galaxies is broader. The right panel shows
how the width of the $\Sg$ distribution grows with increasing kernel
width as the convolution samples coherent structures on larger scales.
To assign significance to the overdensities in galaxies,
we rescale $\Sg$, dividing by its standard deviation.

Next, we remove obvious false positives by rejecting all objects
within the region marked by dashed lines in
Figure~\ref{sgalversussstar}, namely the intersection of the regions
$\Ss < 20$ and $\Sg > 2$.  This removes most, but not all, the false
positives caused by galaxy clusters. Additionally, there remains
contamination from galaxies with large angular size. The SDSS
photometric pipeline mis-classifies HII regions and stellar clusters
in these galaxies as stars. We remove the contaminants by
cross-correlating with the positions of galaxies in the Third
Reference Catalogue (RC3) of \citet{rc3}. Even so, there still remain
objects at a moderately high level of significance whose nature is
unclear. Most of these are probably caused by galaxy clusters or
photometry artifacts, as judged from examination of Hess diagrams and
SDSS image cut-outs, but there may still be a very small number of
genuine Milky Way satellites.

We detect all the known Milky Way satellites, except Boo~II, in a
catalog with magnitude limit $r < 22.5$, analyzed using a kernel with
$\sigma_1=4'$. The most marginal detections are Leo~IV and Ursa
Major~I, which have significances $\Ss = 6.10$ and $5.95$
respectively.  Objects above the threshold are listed in
Table~\ref{tab:zzz}, and include three likely false positives, which
are ``Candidates'' X, Y and Z. The Hess diagrams of these three
detections are shown in Figure~\ref{hess}. The Hess diagrams offer
little evidence to support identification of the candidates as genuine
satellites. Deeper data are needed to provide definitive
classification of the candidates but for the purpose of determining
the satellite luminosity function we exclude all three candidates, as
false positives, from further consideration.

It is prudent to search for candidate satellites on the map convolved
with different inner kernels, since the kernel biases the
algorithm towards objects of a preferred size.  Therefore, we
performed a search on the map convolved with kernels of $2'$ and
$8'$. In the former case, setting the significance to $\Ss > 6.5$
results in the detection of all objects except UMa I and no false
positives; in the latter case, setting $\Ss > 6.0$ includes all objects
except CVn II, Leo IV, LeoT, UMa I and no false positives.

Boo~II, found by \cite{Wa07}, is problematic for our algorithm. Boo~II
contains a very sparsely populated giant branch, and so the brightest
stars are sub-giants and turn-off stars at colors of $g-r <0.5$. Given
our preferred cuts, Boo~II is undetected. It can nonetheless be found
with our algorithm, but only by optimizing the color and magnitude
cuts, for example, to $g-r < 0.5$ and $21 < r <23$.

\begin{figure*}
\begin{center}
 \includegraphics[height=5cm]{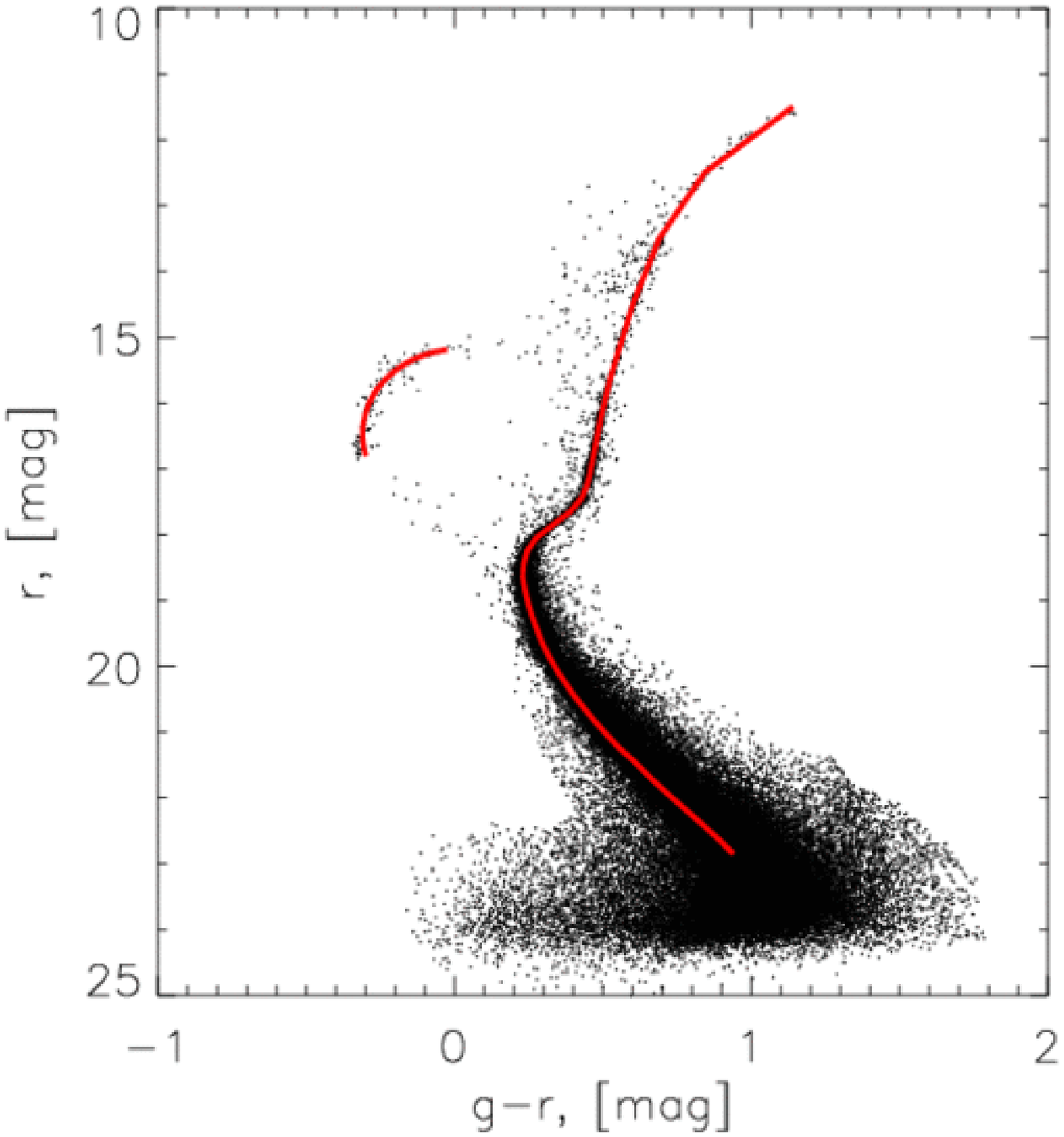}
 \includegraphics[height=5cm]{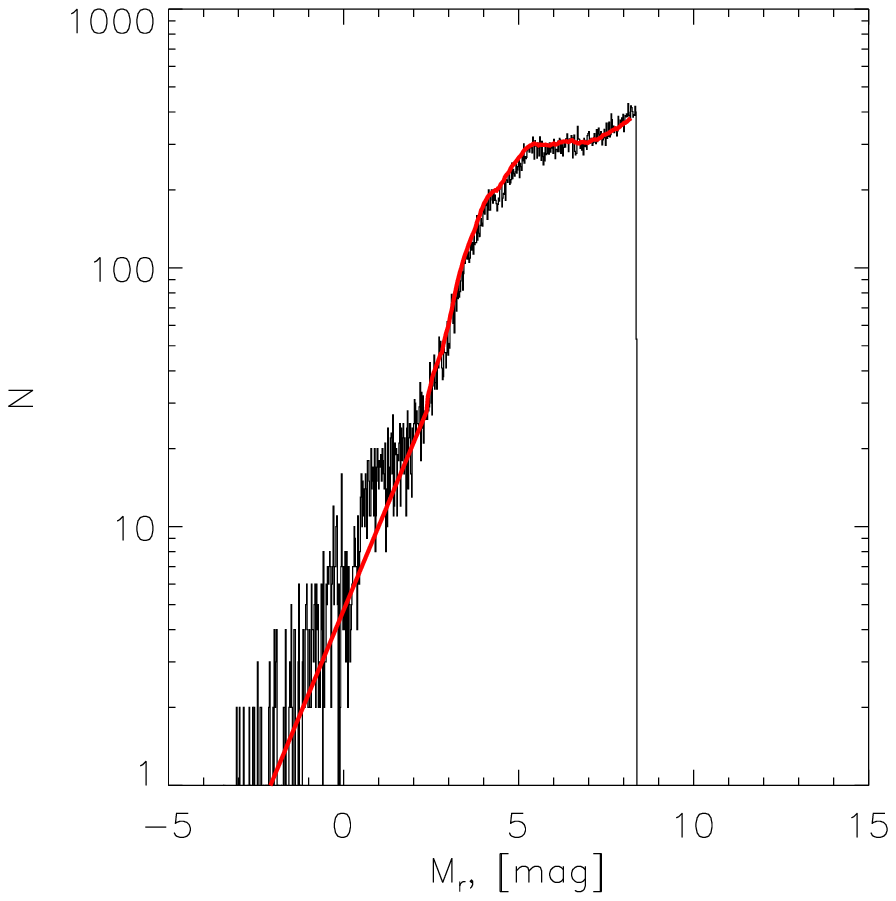}
 \includegraphics[height=5cm]{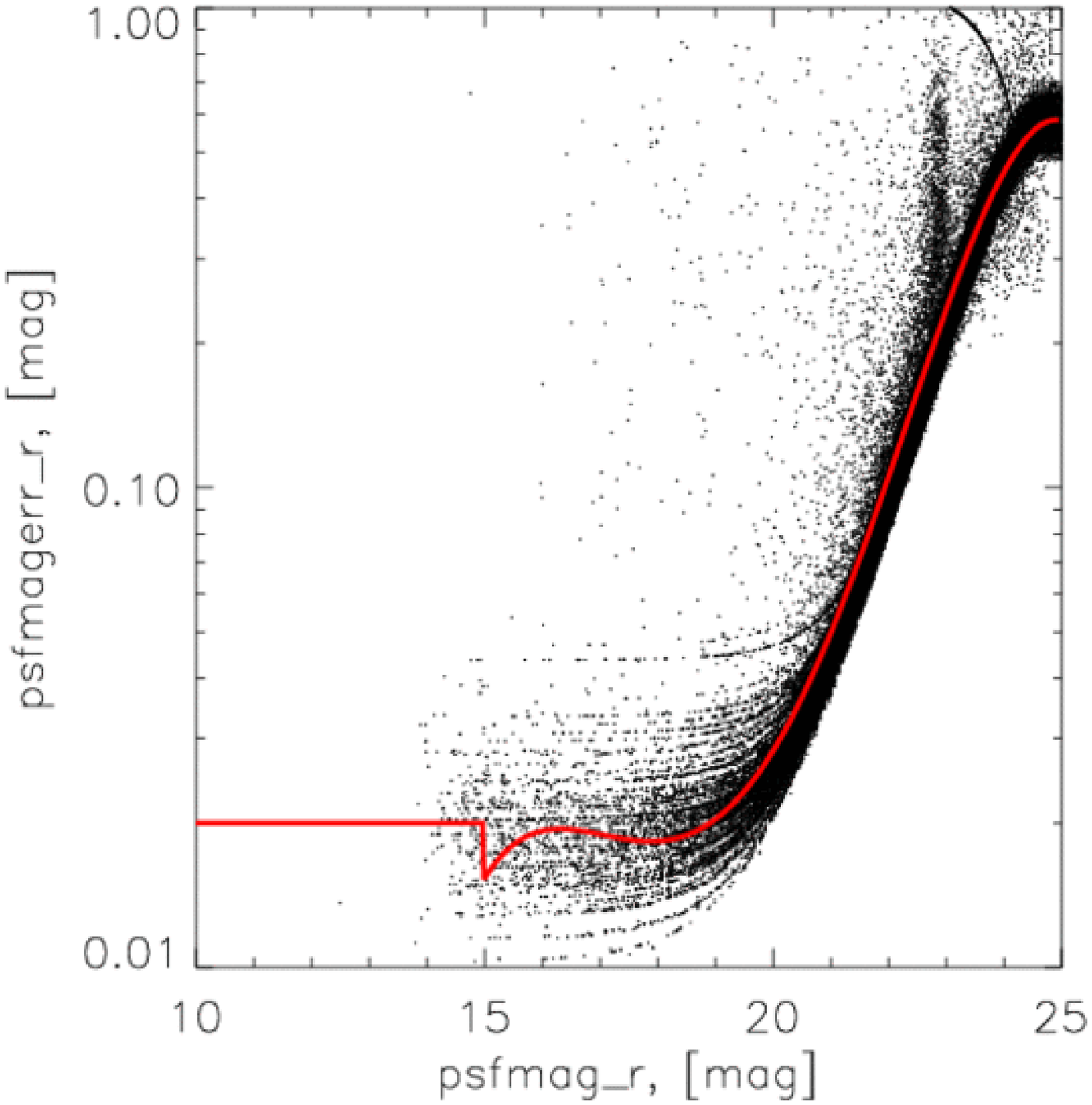}
\end{center}
\caption{Left: M92 color-magnitude data from \citet{clem} used as a
  template for our simulated Milky Way satellites, together with the ridge line
  for the main sequence and red giant branch. The ridge line for the
  horizontal branch is our fit to Clem's (2006) data. Center: The
  observed luminosity function of main-sequence and red
  giant branch stars in M92, together with our model fit of the luminosity 
  function used
  in the simulations. Right: The photometric errors of the SDSS
  $r$-band photometry and our model fit used in the simulations}
 \label{m92_lmf}
\end{figure*}
\begin{figure*}
\begin{center}
 \includegraphics[width=15cm]{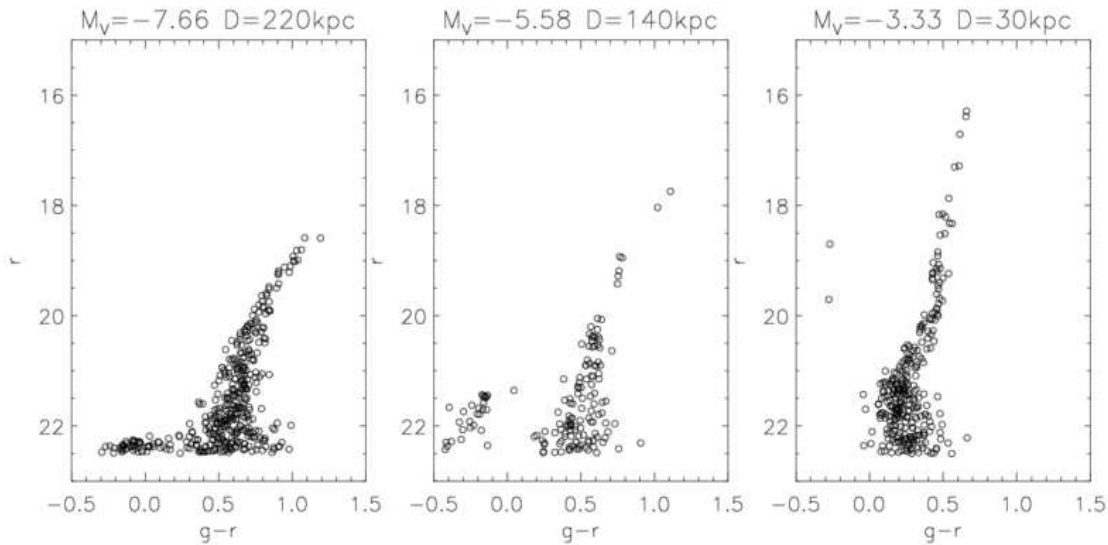}
\end{center}
 \caption{Simulated color-magnitude diagrams for \change{hypothetical}
   Milky Way satellite galaxies with properties close to those of
   Canes Venatici~I, Hercules and Ursa Major~II---the actual
   color-magnitude diagrams of these galaxies are given in
   \citet{Zu06a,Zu06b} and \cite{Be07}.}
 \label{simul_show}
\end{figure*}
\begin{figure*}
\begin{center}
 \includegraphics[width=19cm]{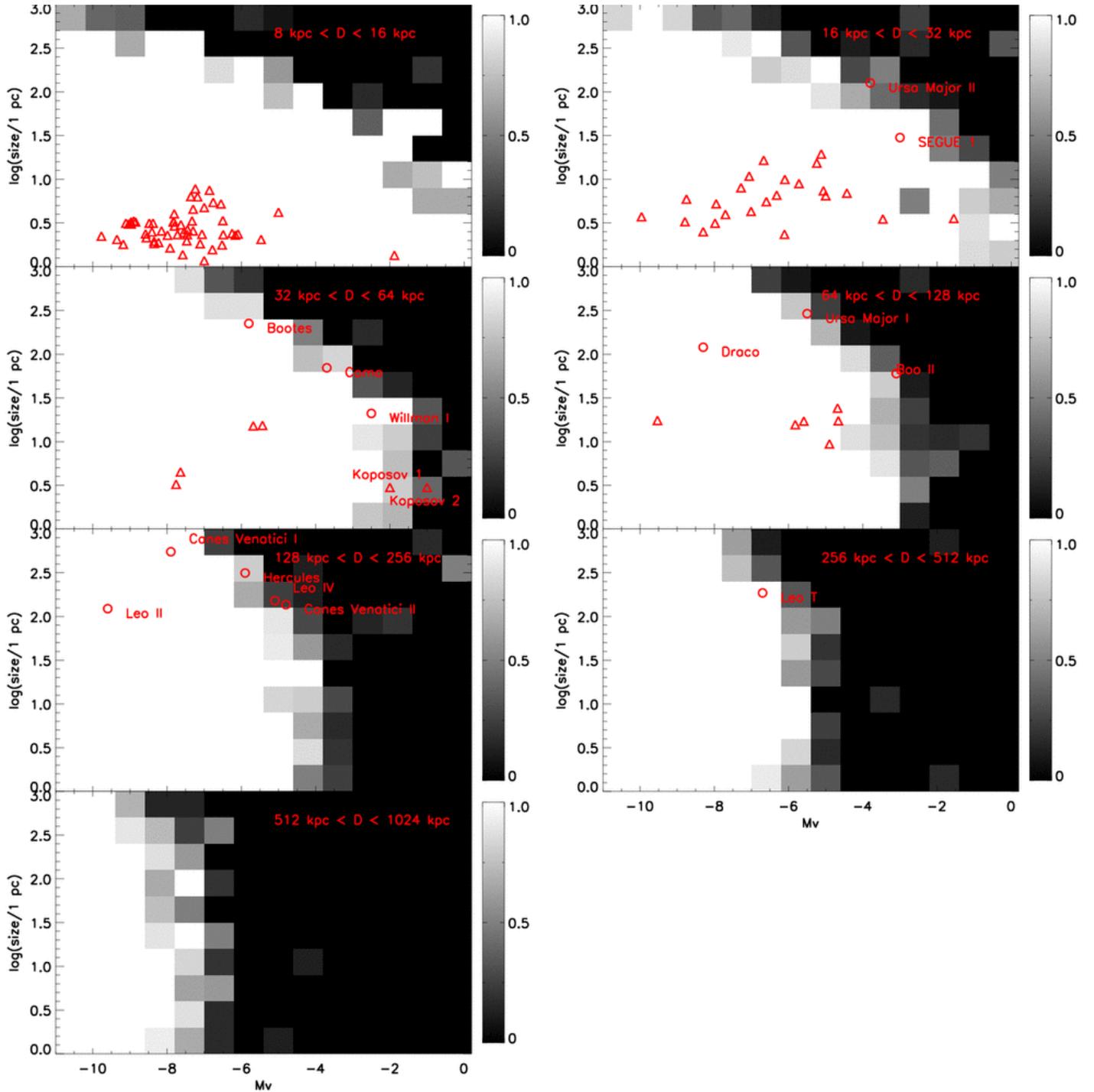}
\end{center}
 \caption{Detection efficiency maps for Milky Way satellites, shown as a
   function of luminosity and size for different distance bins. White indicates
   100\% detection efficiency, black indicates 0\%. Red circles mark the
   locations of the known dwarf galaxies, red triangles the known globular
   clusters (data taken from~\citet{Ha96}). Notice that many of
   the very recent SDSS satellite galaxy discoveries occur near the
   boundary, where the detection efficiency is changing rapidly.}
 \label{efficiencies_known}
\end{figure*}
\begin{figure}
\begin{center}
 \includegraphics[height=7cm]{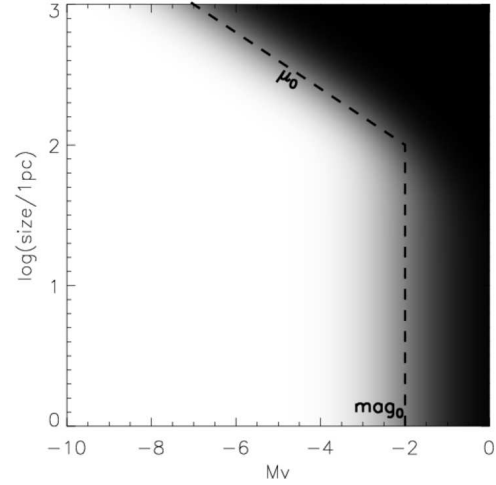}
\end{center}
\caption{Characterizing the satellite galaxy detectability: this illustrative
  figure shows the model function $\epsilon(M_V,\mu)$ (from
  Eq.~\ref{eq:efficiency_model}) used to fit the observed detection
  efficiencies from the simulations and demonstrating the role played
  by the thresholds $M_{V,{\rm lim}}$ and $\mu_{\rm lim}$. The
  function parameters used to produce the plot were 
  $M_{V,{\rm lim}}=-2\,{\rm mag}$,
  $\mu_{\rm lim}=29.5\,{\rm mag\, arcsec^{-2}}$, $\sigma_M=1\,{\rm mag}$,
$\sigma_\mu=1\,{\rm mag}$}
 \label{efficiency_model}
\end{figure}
\begin{figure*}
\includegraphics[width=16cm]{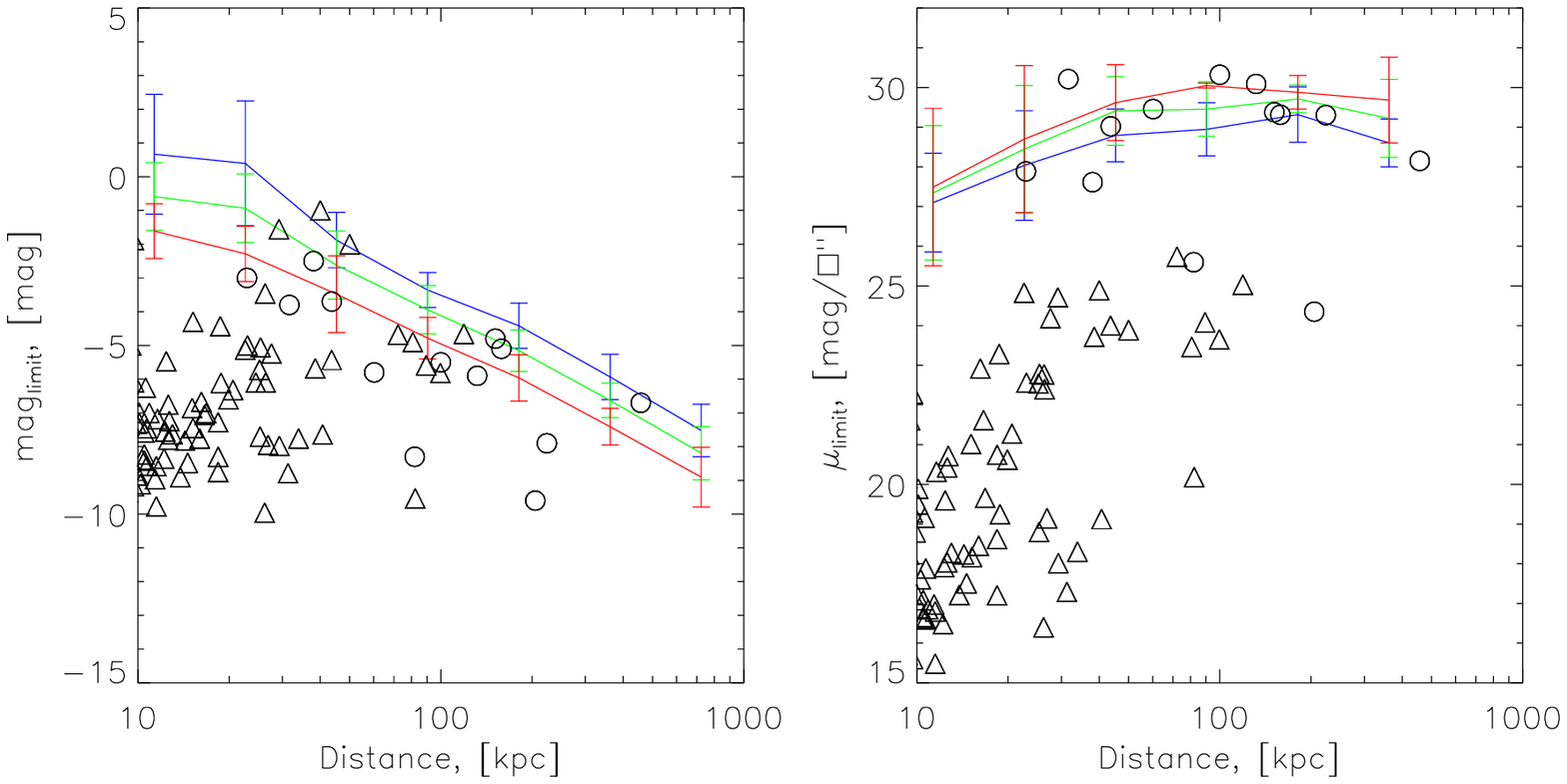}
\includegraphics[width=16cm]{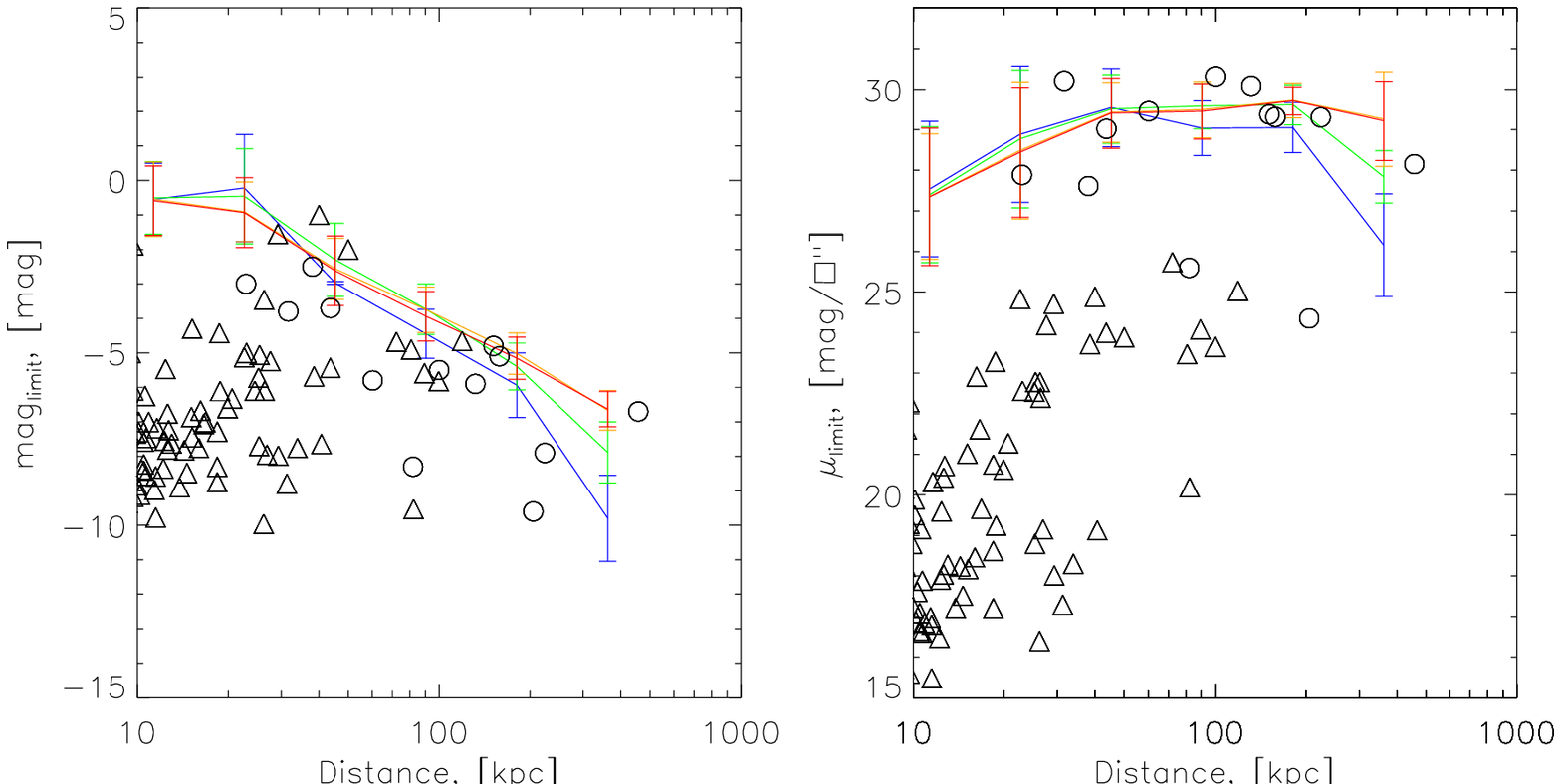}
\caption{The known satellites and globular clusters shown in two-dimensional
plots of Galactocentric distance versus absolute magnitude (left) and surface
brightness (right).  Circles mark the locations of the known dwarf galaxies,
triangles the globular clusters.  The error bars show either $\sigma_M$ or
$\sigma_\mu$ derived from our model fits (see Eq.~\ref{eq:efficiency_model}).
The detectability of the objects depends on their location relative to the
limiting absolute magnitude (left) and surface brightness (right) as a function
of Galactocentric distance for each kernel size/color cut employed in the
search.  Upper panels:  The three lines show the detection limits for different
sizes of the inner Gaussian in the kernel (blue -- 2\arcmin, green -- 4\arcmin,
red -- 8\arcmin). Lower panels: The four lines show the detection limits for
the different $g-r$ color cuts employed (black -- 0.2, blue --
0.4, green -- 0.6, orange -- 0.9, red --1.2) and fixed kernel size of 4\arcmin. 
}
 \label{mag_sb_vs_distance_kernel}
\end{figure*}

\section{Application to Simulated Data}
To test our detection algorithm, we carry out an extensive set of
simulations in which we add mock dwarf galaxy satellites and globular
clusters to the SDSS DR5 catalog.  In particular, we add to the
\change{catalog the} $g$ and $r$ magnitudes of stars from the
simulated objects, at specified right ascensions and declinations.
These augmented catalogs are then fed through our automated pipeline,
and the number of stellar overdensities with significance above the
threshold is calculated as a function of distance, size and
luminosity.  We explore how changes in the $g-r$ color cuts and kernel
sizes ($\sigma_1$ from Eq.~\ref{eq:sigmaL}) affect the efficiency of
the algorithm.

The $g$ and $r$ photometry of all simulated objects is based on that
of the globular cluster M92.  The left panel of Figure~\ref{m92_lmf}
shows the color-magnitude diagram (CMD) of M92, together with a
main-sequence and red giant branch ridgeline from~\citet{clem}, to
which we have added a horizontal branch ridgeline. From the $r$-band
data, we construct a main-sequence and red giant branch luminosity
function and approximate it with a smooth fit, as shown in the middle
panel of Figure~\ref{m92_lmf}. We also determine the luminosity
function for the stars on the horizontal branch ridgeline. We populate
the ridgelines using the luminosity function.  The choice is appropriate,
as M92 (12 Gyrs, [Fe/H] $\approx -2$) is typical of the old,
metal-poor populations in the Milky Way
satellites~\citep[see e.g.][]{van00}. 
Additionally, we add a scatter in $r$- and $g$-
magnitudes, derived from a fit to the errors in the SDSS point-spread
function photometry, as illustrated in the right panel of
Figure~\ref{m92_lmf}.

The spatial distribution of stars in the simulated objects is chosen
to follow a Plummer law, which is a reasonable fit to
most of the Milky Way dwarf
spheroidals~\citep[see e.g. ][]{Ir95,Kl02}. 
For ellipticities less than 0.5 -- which
corresponds to the most flattened of the SDSS discoveries, Hercules
and Ursa Major~II~\citep{Zu06b,Be07} -- the detection efficiency of
objects barely changes with ellipticity. The Plummer radius,
luminosity and distance are chosen to cover uniformly in logarithmic
space the following ranges: Plummer radius 1\,pc $<r_h<$ 1\,kpc,
luminosity $-11 \lesssim M_v \lesssim 0$ and heliocentric distance
10\,kpc $< D <$ 1\,Mpc.  We generate 8000 galaxies with random right
ascension and declination within the DR5 footprint.  We then split the
simulated sample into 20 distance bins to eliminate overlap
between simulated objects. The stars from the simulated galaxies are
added to the DR5 stellar catalog.  Figure~\ref{simul_show} shows mock
CMDs for simulated objects matching the recently discovered dwarf
galaxies Canes Venatici I, Hercules and Ursa Major
II~\citep{Zu06a,Be07,Zu06b}.  These are good approximations
to the observed CMDs of these objects.

In our simulations, we test several inner kernel sizes. The reason is
that for a given distance, the kernel size gives rise to an optimum
physical size of the detectable objects. For example, at a distance
of 50\,kpc, a kernel size of $4'$ corresponds to a physical size of
$\simeq$60\,pc.  As we want our algorithm to be sensitive to objects
of different sizes, we use three different inner kernel sizes, namely
$\sigma_1 = 2'$, $4'$ and $8'$. An object is considered to be detected
if it is above a threshold on the map convolved with at least one of the
kernels (the threshold for the $2'$ kernel is 6.50, for the $4'$ kernel -- 5.95
and for the $8'$ kernel -- 6.00, see Section~\ref{observ_section}).
We refer to this procedure as the combined kernel. This is equivalent
to the algorithm used in the previous Section, because the list of
detections for $2'$, $4'$ and $8'$ kernels includes all the known
dwarfs.

Figure~\ref{efficiencies_known} shows two-dimensional efficiency maps
as a function of luminosity and size in seven distance bins spanning
the range 8\,kpc to 1\,Mpc. For Figure~\ref{efficiencies_known}, we
have used the color cut of $g-r < 1.2$ and the combined kernel,
together with an outer kernel of size $\sigma_2 =60'$.  Black
corresponds to zero detection efficiency, and white to unit
efficiency.  The locations of the known Milky Way globular clusters
and dwarf galaxies in this parameter space are recorded as red
triangles and circles. While a number of known objects lie well within
the efficiency boundary, some of the recent discoveries lie close the
boundary.  It is evident that there is no steady gradient in
efficiency, but rather a steep boundary between detectability and
non-detectability.  In fact, the primary contribution to the
finite-extent of the gradient visible in the Figures is produced by
the significant extent of the individual distance bins (the width of
the distance bins is 0.3 dex). The pixel size in magnitude is 0.8, and
in $\log \rh$, it is 0.3. This means that there are typically 10
objects in each bin and so we expect moderate fluctuations due to shot
noise.

As the efficiency changes so quickly near the boundary, and as several
objects lie close to this zone, we carried out more detailed
simulations on objects similar to the known dwarfs. We created 1000
Monte Carlo realizations of each of the known dwarfs, and fed them
into the pipeline. Table~\ref{tab:effs_simul} lists the derived
detection efficiencies for each object. The detection efficiencies are
$\gtrsim$ 50 \%, with the sole exception of Boo~II, confirming our
assertion that the known satellites possess high detection
probabilities.

For the regime in which objects are larger than the kernel size, some
of the stars belonging to the satellite are missed by the window
function, and for such objects the detectability is determined by the
number of stars within the window function, i.e. the surface
brightness. This effect produces the surface brightness limit seen in
Figure~\ref{efficiencies_known}.  For objects smaller than the kernel
size, all the stars are within the window function regardless of the
size of the objects, therefore for such objects, the detectability
doesn't depend on physical size, but depends only on the total number
of stars, i.e. the luminosity. \change{This} effect produces the rapid
change in detection efficiency at fixed absolute magnitude evident in
Figure~\ref{efficiencies_known}.
These two regimes can be modeled with thresholds in surface brightness
and absolute magnitude by adopting a functional form:
\begin{equation}
\epsilon (M_v, \mu) =  G \left( \frac{M_V - M_{V,{\rm
      lim}}}{\sigma_M} \right)   
G \left( \frac{\mu - \mu_{\rm lim}}{\sigma_\mu} \right),
\label{eq:efficiency_model}
\end{equation}
where $G$ denotes the Gaussian integral, which is defined as
\begin{equation}
G(x) = {\frac{1}{\sqrt{2\pi}}} \int_{x}^\infty \exp(-t^2/2) dt. = \frac{1}{2}
\,erfc\left(\frac{x}{\sqrt 2}\right)
\end{equation}
To describe the detectability in each distance bin, there are four
parameters that are fitted -- namely the detection thresholds in
surface brightness $\mu_{\rm lim}$ and absolute magnitude 
$M_{V,{\rm lim}}$, together with their widths $\sigma_\mu$ and $\sigma_M$.
As an illustrative example, the grey-scale map of the efficiency
function $\epsilon(M_{V},\mu)$ from~Eq.~\ref{eq:efficiency_model} is
shown in Figure~\ref{efficiency_model}, with dashed lines indicating
the thresholds. Note the shape of the detection boundary, with the
prominent ``knee'', which corresponds to the boundary between the two
detection regimes for objects of different sizes at fixed distance, as
described above.

The two key parameters for the detection pipeline are the inner
kernel size $\sigma_1$ and the color cut applied to the source
catalogs. The top two panels of
Figure~\ref{mag_sb_vs_distance_kernel} show the dependence of
$M_{V,{\rm lim}}$ and $\mu_{\rm lim}$ on distance, when convolved with
the three different inner kernels. For a
given kernel, the limiting magnitude declines roughly linearly with
the logarithm of distance. Objects at the limiting magnitude have an
apparent size that is smaller than the kernel size and their
detection significance is reduced by the background
contribution. Shrinking the kernel size removes some of the background
and increases the significance of fainter satellites. The dependence of the
$M_{V,{\rm lim}}$ and $\mu_{\rm lim}$ on distance
for the combined kernel is not plotted, because in the top left panel of
Figure~\ref{mag_sb_vs_distance_kernel} the combined kernel basically follow the
dependence of $2\arcmin$ kernel and in the top right panel
of Figure~\ref{mag_sb_vs_distance_kernel} the combined kernel follow
the dependence of $8\arcmin$ kernel.
This is illustrated in the top left panel of
Figure~\ref{mag_sb_vs_distance_kernel}, where it is clear that a
smaller kernel allows us to detect fainter objects. However, as the
top right panel shows, this is at the expense of satellite size.
Larger kernels pick up more stars from extended objects and hence
reach fainter surface brightness. 
When combining different kernels in the pipeline, the overall
    limits in surface brightness and absolute magnitude (2', 4',  
    8', see Section~\ref{observ_section}) can be approximated by  
    the blue line in the top left panel of
    Figure~\ref{mag_sb_vs_distance_kernel} and the red line in the
    top right panel of Figure~\ref{mag_sb_vs_distance_kernel}.
    Smaller kernels allow the detection of galaxies that are low
    in absolute magnitude, and larger kernels allow the detection
    of galaxies that are fainter in surface brightness.
It is also reassuring to see that the error bars $\sigma_M$ are of the same
order as the difference in the limiting magnitude moving to a neighboring
bin.

We explore the effects of changing color cuts and report the results
in the bottom two panels of
Figure~\ref{mag_sb_vs_distance_kernel}. The color cut of $g-r < 0.4$
can improve slightly the magnitude limit for nearby objects by
selecting turn-off stars. This improvement deteriorates rapidly as we
exhaust the supply of turn-off stars. At larger distances, red color
cuts like $g-r < 1.2$ are more efficient at picking up giant
stars. The same effect explains the drop in $\mu_{\rm lim}$ and
$M_{V,{\rm lim}}$ at large distances for bluish color cuts. Our choice
of color cut $g-r < 1.2$ is conservative, mostly eliminating thin disk
stars, and is, overall, the best-behaved and most robust. It also allows
us to minimize the influence of metallicity and age changes in the
stellar population of the satellites.

\begin{deluxetable}{cc}
\tablecaption{Detection Efficiencies of Simulated Objects
Resembling Known Satellites.\label{tab:effs_simul}}
\tablehead{ \colhead{Object} & \colhead{Efficiency}}
\startdata
Bootes & 1.00\\
Draco & 1.00\\
Leo I & 1.00\\
Leo II & 1.00 \\
SEGUE 1 & 1.00 \\
Canes Venatici I& 0.99 \\
Willman I & 0.99\\
Coma & 0.97\\
Koposov 1 & 0.90\\
Leo IV & 0.79 \\
Ursa Major II & 0.78\\ 
Leo T & 0.76\\
Hercules & 0.72\\
Ursa Major I & 0.56\\
Koposov 2 & 0.48 \\
Canes Venatici II & 0.47\\
Boo II & 0.20\\
\enddata
\end{deluxetable}
\begin{deluxetable}{ccc}
\tablecaption{Limiting Satellite Absolute Magnitude and Surface
Brightness as a Function of Distance 
}
\tablehead{
\colhead{Distance} & 
\colhead{$M_{V,lim}$} &
\colhead{$\mu_{lim}$}
\\
\colhead{kpc} &
\colhead{} &
\colhead{$\rm mag/\square''$}
}
\startdata
11 &  0.6 & 27.5\\
22 &  0.4 & 28.7\\
45 &  -1.9 & 29.6\\
90 &  -3.4 & 30.0\\  
180 & -4.4 & 29.9\\
260 & -5.9 & 29.9\\
720 & -7.5 & 29.6
\enddata
\end{deluxetable}
\begin{figure}
 \plotone{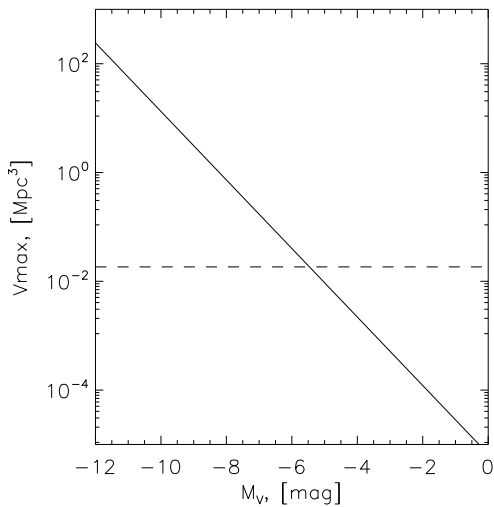} 
 \caption{The accessible volume within \change{the} DR5 footprint for
   galaxies with different luminosities and surface brightnesses
   $\mu_{\rm lim}$, $\mu \lesssim 30\,mag/\Box''$
   (see Figure~\ref{mag_sb_vs_distance_kernel}. The volume limited by the
   virial radius (280\,kpc) and within DR5 is shown by the dashed line. 
}
\label{vmax}
\end{figure}
%
\begin{figure}
 \plotone{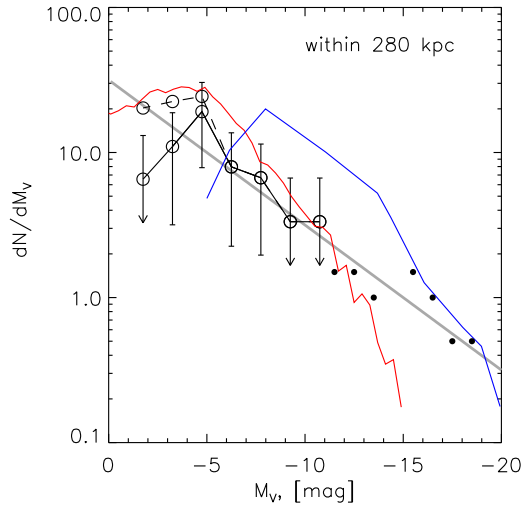}
 \caption{The luminosity functions of Milky Way satellite galaxies within
   $\sim$280\,kpc (virial radius) inferred from our analysis under the
   assumption of two different radial distributions of satellites, NFW-like
   (solid black line) and isothermal (dashed black line). The
   calculation uses the satellite list
   and the volume correction factor obtained with the pipeline using
   the cuts $r<22.5$ and $\Ss >5.95$. The arrows on error
   bars indicate that there is only one galaxy in the particular bin,
   and so the Poisson error is formally 100\%.
  The theoretical prediction of
   Figure~1 of \citet{Be02} is shown in a red line, and the prediction
   of \citet{So02} for $z_{reion}=10$ is shown as a blue
   line. Additionally, the luminosity function for the bright
   ($M_V<-11$) satellites of the Milky Way sampled over the
     whole sky together with the bright M31 satellites within 280\,kpc
   from \citet{Me07} is plotted with filled small symbols (the list of
   plotted objects consists of Sgr, LMC, SMC, Scu, For, LeoII, LeoI,
   M32, NGC 205, And I, NGC 147, And II, NGC 185, And VII, IC 10).
   The function $dN/dM_V=10\times 10^{0.1\,(M_V+5)}$ is shown in grey.
 }
 \label{satdist_models}
\end{figure}

\section{The Luminosity Function}

\subsection{Analysis of the detection efficiency maps}

With \change{an understanding of which} satellite galaxies can be
detected in SDSS DR5, together \change{with our sample} of actual
detections, we can now estimate the luminosity function of faint Milky
Way satellites. We start by re-examining the efficiency maps in
Figure~\ref{efficiencies_known}, where the locations of the known
Milky Way globular clusters and dwarf galaxies are overplotted as red
triangles and circles respectively.  We can conclude that within the
DR5 footprint there are certainly no bright satellites (either
globulars or galaxies) nearby ($D < 32$ kpc) that have eluded
discovery. However, the disrupting galaxy UMa~II~\citep{Zu06b}
provides a clue as to the likely locations of remnants. It is still
possible that disrupted galaxies remain undiscovered nearby. They can
lurk in the black portions of the uppermost two panels of
Figure~\ref{efficiencies_known}.

All that has survived in the inner Galaxy ($8 < D< 16$ kpc) is a
population of globular clusters, which occupy a small region in the
luminosity and size parameter space. They are predominantly old
globular clusters belonging to the bulge. Only the densest survive
against the disruptive effects of Galactic tides and shocking, which
is illustrated by the apparent size bias. Notice that the datapoints
lie well away from the detection boundary, suggesting that the sample
is complete at least within $8 < D< 16$. Moving outwards ($16 <D< 32$
kpc), the globular clusters belong to the halo and may have been
accreted \citep{Mackey04}.  Their size distribution is broader. Some
of these objects are in the process of disruption, such as Pal 5 and
NGC 5466 ~\citep{Od01,De04,Be06c,Fe07}. The very faint and distant
globular clusters discovered recently by \citet{Ko07} are visible in
the third panel of Figure~\ref{efficiencies_known} ($32 < D< 64$ kpc)
right on the border of detectability. Further such sparse globulars
may remain undetected.

Beyond 30\,kpc, the dwarf spheroidals begin to appear.  The
long-known dwarfs such as Draco and Sculptor lie far from the
boundary, in regions of the luminosity and size parameter space where
the DR5 search efficiency is unity.  However, all the recent SDSS
discoveries, such as Canes Venatici I, Bootes and Hercules, lie close
to the detection boundary, where the efficiency declines rapidly
from unity to zero.  \citet{Be07} claimed that there is a paucity of
objects with half-light radii between $\sim 40$\,pc and $\sim 100$\,pc.
Our calculations \change{support the idea that the gap is real and not
produced by selection effects}.  If there were objects with radii
between $\sim 40$\,pc and $\sim 100$\,pc, there is a broad range of
parameter space in which they would have been found.

Although most of the new detections lie in the gray areas of the plot,
the empty white regions with unit efficiency are telling us something
important.  There are swathes of the parameter space in which
we would have detected objects if they existed. For example,
there are very few bright objects ($M_v < -6$). The absence of
detections of bright objects does by itself provide a strong
constraint on the luminosity function of Milky Way satellites. There
also do not appear to be any analogues of the extended, luminous star
clusters found in M31 by \cite{Hu05}.  Although SDSS data may still
contain evidence for further, hitherto unknown, dwarf galaxies, it is
unlikely that their nature can be unambiguously established without
substantial quantities of follow-up imaging.
\change{We emphasize that, since we never probe fainter than a
  certain surface brightness limit, an even
  larger population of very low surface brightness galaxies -- which can
  not be detected with SDSS -- may exist.}

\subsection{Estimation of the Luminosity function}

Figure~\ref{vmax} shows the accessible volume for galaxies of
different luminosities probed by our algorithm (which in practice is a
function mostly of the luminosity) within the SDSS DR5 footprint. As the
logarithm of distance scales roughly linearly with limiting magnitude
(see Figure~\ref{mag_sb_vs_distance_kernel}), so does the logarithm of
the accessible volume. Using this, and the fact that the SDSS survey
covers $\sim 1/5$ of the sky, we can convert the set of known objects
into a volume corrected luminosity function\footnote{The existing data
  on the globular cluster population indicate that at least some
  globular clusters have complicated metallicity, age distributions
  and kinematics and may in fact be stripped nuclei of dwarf galaxies
  \citep{Zi88,Pi07}. \change{The selection of such objects which are
    considered as dwarf galaxies is an additional source of
    uncertainty in any luminosity function determination.}}.

The observed luminosity function is constructed using all the
well-established dwarf galaxies in DR5, namely Leo II, Draco, Leo I,
CVn I, Boo I, Hercules, UMa II, Com, CVn II, Leo T, UMa~I, Leo~IV as
well as the possible dwarf Willman 1. Segue 1 is not used because it
is not in DR5~\citep{Be07}, and Boo~II is not used because it is not
detected with our adopted identification-pipeline parameters. All the
satellites included in our calculation have a surface brightness of at
least $30$ mag arcsec$^{-2}$. To relate the observed number of
satellite galaxies in our sample to the total number of satellites in
the Milky Way halo, it is necessary to adopt an underlying radial
distribution of satellite galaxies (see Appendix).  In a given
magnitude interval, we know the observed number of satellites within
$V_{\rm max} (M_v)$ from Figure~\ref{vmax}, together with their
detection efficiencies from Table~\ref{tab:effs_simul}. If we assume a
number density law $n(r)$ for the satellites, then its normalization
at each magnitude interval can be fixed by integrating the density law
out to $V_{\rm max}$.  The total number of satellites within 280\,kpc
(the virial radius of the halo)
is now the integral of the density law out to this limit.
Figure~\ref{satdist_models} shows the results of the
calculation for two such density laws.
The dashed line shows the luminosity
function assuming the satellites are distributed in an isothermal
sphere (namely, $n(r) \propto 1/r^2$). The solid line shows the
luminosity function if the density fall-off is steeper at large radii 
($n(r) \propto 1/r^3$, like Navarro-Frenk-White profile, although to prevent
the $1/r^3$ profile from diverging in the MW center we use $n(r) \propto
r^{-2}(r+r_{\rm c})^{-1}$ with the core radius $r_c = 10$\,kpc).  Of
course, the nature of some of the objects we have \change{included} in
the dwarf galaxy luminosity function is still uncertain -- in
particular, Willman 1 may be a globular cluster, although
\citet{Martin07} provide evidence for a metallicity spread and dark
matter content.  It is unclear whether Leo T should be included or
excluded, as it is most likely a transition object with rather
different properties from the other dwarf spheroidal galaxies in our
sample.  The error bars in Figure~\ref{satdist_models} are given by
the square root of the number of datapoints in the absolute magnitude
interval divided by the volume correction factor. At the bright end,
the error bars are large, since we have only two objects with $M_v<-9$,
namely Leo~I ($M_v= -11.5$) and Leo~II ($M_v=-9.6$). At the faint end,
the error bars are also large because of the substantial volume
correction factor.
In Figure~\ref{satdist_models}, we show the luminosity function for
satellites within 280\,kpc (a proxy for a Milky Way virial radius
\citep{Kly02,Be02}). To define the bright end of the luminosity
function, which cannot be reliably determined from our data
\change{since DR5 does} not contain dwarfs brighter than $M_V \sim
-11$, we have also included in Figure~\ref{satdist_models} the
estimate of the luminosity function (filled points) for the bright
satellites of the Milky Way sampled over \change{the full sky},
together with the bright M31 satellites within 280\,kpc from
\citet{Me07}.  In Figure~\ref{satdist_models}, we also overplot the
power-law function $dN/dM_V=10\times 10^{0.1\, (M_V+5)}$, which
approximates the datapoints in the range of $-19<M_V<-2$ (with
probably some flattening at $M_V\sim -4$). The integration of this
power-law gives approximately 45 dwarfs brighter than -5.0, and 85
dwarfs brighter than -2.0.


There are a number of theoretical predictions of the luminosity
function of the Local Group in the literature. For example,
\cite{So02} shows the results of semi-analytic galaxy formation
calculations, including the effects of supernova feedback and
photoionization. The luminosity function from \citet{So02} for
$z_{reion}=10$ \citep{Pa07} are plotted with blue line in
Figure~\ref{satdist_models}.  Although the numbers
of luminous satellites are in reasonable agreement with the data, the
shape of the luminosity function is not. All Somerville's (2002)
luminosity functions turn over at $M_v \approx -9$ or brighter,
depending on the epoch of reionization, whereas the luminosity
function derived in Figure~\ref{satdist_models} turns
over fainter than $M_V \approx -5$, if at all. Therefore,
Somerville's (2002) theoretical calculations overproduce Draco-like
objects ($M_V \approx -10$) by a factor of a few, and underproduce
much fainter galaxies like Boo ($M_V \approx -6$) by almost an order
of magnitude.

\cite{Be02} also provides calculations of the luminosity function of
the Milky Way satellites, including the effects of tidal disruption as
well as photoionization. They report the luminosity functions for
dwarfs with a range of of different central surface brightness cuts,
namely $18, 20, 22, 24$ and $26$ mag arcsec$^{-2}$, the last of which
is plotted in Figure~\ref{satdist_models} in a red line.  At first
glance, the fit seems plausible, especially given the size of the
error bars on the datapoints. The turn over in Benson et al.'s
luminosity function is at $M_V \approx -3$ and the numbers of
predicted satellites at faint magnitudes are also consistent given the
uncertainties. However, Benson et al.'s model significantly
underproduces the number of bright satellites. Additionally, Benson et
al.'s satellites have a much higher central surface brightness -- our
SDSS survey corresponds to a surface brightness cut of $\sim 30$ mag
arcsec$^{-2}$.  Figure 2 of \citet{Be02} does show the luminosity
function for all objects, irrespective of surface brightness.
Although there has been a large change in the luminosity function on
moving from a detection threshold of $22$ to $26$ mag arcsec$^{-2}$, 
there is only a small change on moving from $26$ to $\infty$ mag 
arcsec$^{-2}$.

\section{Conclusions}

There have been persistent discrepancies between the observed numbers
of Milky Way satellites and the predictions from numerical simulations
of galaxy formation for a number of years.  Although here has been a
cavalcade of discoveries of new Milky Way satellites using the SDSS
over the last two years, a systematic search -- with quantifiable
detection limits and efficiencies -- not been undertaken.  In this
paper we have presented a quantitative search methodology for Milky
Way satellite galaxies in SDSS data and have used this method to
\change{compute} detection efficiency maps, which ultimately allow the
construction of the satellite galaxy luminosity function.

In our method, the star count map is convolved with a family of
kernels which are the difference of two Gaussians.  Intuitively, this
algorithm can be understood as constructing an estimate of the local
stellar density minus the background. By attaching a statistical
significance to the overdensities in the convolved image, this enables
us to construct a ranked list of candidates.  Although this idea is
simple enough, its practical application is hampered by the fact that
the separation between stars and galaxies by the SDSS pipeline becomes
unreliable at magnitudes fainter than $r \simeq 22.5$. The resulting
false positives must be removed by cross-correlating with galaxy
catalogs.  The significance threshold of peaks in our survey is set by
requiring the detection pipeline to produce a ``clean'' list of Milky
Way satellites.

To compute the detection efficiency, we create mock SDSS catalogs with
stars from simulated dwarf galaxies and use Monte Carlo methods to estimate
recovery as a function of satellite galaxy parameters and heliocentric
distance. There is a sharp boundary between detectability and
non-detectability. The efficiency maps make clear that there are
large domains in parameter space in which objects would have been
detected had they existed. In particular, even at heliocentric
distances as great as 1\,Mpc, objects brighter than $M_v \sim -8$ would
have been detectable in SDSS. Similarly, populations of extended,
luminous star clusters would have been found in SDSS, if they existed
in the Milky Way.

With the efficiency in hand, we can -- for the first time -- correct
the observed luminosity function of the Milky Way satellites for
selection effects and compute the true luminosity function.  The
number density of satellite galaxies continues to rise well below
$M_V\sim -8^m$; depending on the radial distribution model assumed it
may or may not flatten or turn over at $M_V \gtrsim - 5$.  Overall,
the luminosity function of {\it all} Milky Way satellites may be
reasonably well described by a power-law, $dN/dM_{V}= 10 \times
10^{0.1 (M_V+5)}$ from $M_V=-2$ to -18.  This power-law predicts
$\sim$ 45 satellites brighter than $M_V=-5$, and $\sim$85 satellites
brighter than $M_V=-2$.  The normalization of the luminosity function
is in reasonable agreement with the predictions of semi-analytic
modeling of galaxy formation, but the shape is not.  There also
remains a discrepancy in the distribution of surface brightnesses of
such objects, in the sense that the semi-analytic models underproduce
dwarfs with a central surface brightness fainter than $26$ mag
arcsec$^{-2}$.

\acknowledgments
S.  Koposov is supported by the DFG through SFB 439 and by a EARA-EST Marie
Curie Visiting fellowship.  VB acknowledges the award of a a Postdoctoral
Research Fellowship from the Science and Technology Facilities Council (STFC).
NWE, PCH and DZ acknowledge support from the STFC-funded Galaxy Formation and
Evolution programme at the Institute of Astronomy.  We thank Andrew Benson for
supplying us with numerical data on his semi-analytic calculations.  Funding for
the SDSS and SDSS-II has been provided by the Alfred P.  Sloan Foundation, the
Participating Institutions, the National Science Foundation, the U.S.
Department of Energy, the National Aeronautics and Space Administration, the
Japanese Monbukagakusho, the Max Planck Society, and the Higher Education
Funding Council for England.  The SDSS Web Site is http://www.sdss.org/.

The SDSS is managed by the Astrophysical Research Consortium for the
Participating Institutions. The Participating Institutions are the
American Museum of Natural History, Astrophysical Institute Potsdam,
University of Basel, Cambridge University, Case Western Reserve
University, University of Chicago, Drexel University, Fermilab, the
Institute for Advanced Study, the Japan Participation Group, Johns
Hopkins University, the Joint Institute for Nuclear Astrophysics, the
Kavli Institute for Particle Astrophysics and Cosmology, the Korean
Scientist Group, the Chinese Academy of Sciences (LAMOST), Los Alamos
National Laboratory, the Max-Planck-Institute for Astronomy (MPIA),
the Max-Planck-Institute for Astrophysics (MPA), New Mexico State
University, Ohio State University, University of Pittsburgh,
University of Portsmouth, Princeton University, the United States
Naval Observatory, and the University of Washington. 
This research has made use of the SAI Catalogue Access Services, Sternberg
Astronomical Institute, Moscow, Russia

\appendix
\section{The Calculation of the Correction to the Luminosity Function}

To calculate the luminosity function of Milky Way satellites within
$r_{\rm LF}=280$\,kpc, we select all the satellites within DR5 which
are interior to $r_{\rm LF}$, and construct the histogram of $M_V$ of
these objects. From the simulations, we know that not all objects are
detected with 100\% efficiency and the histogram $h(M_V)$ is weighted with the
object detection efficiencies.
$$h(M_V) = \sum\limits_i \frac{1}{\epsilon_i} \delta(M_V, M_{V,i})$$ where 
$\epsilon_i$ is the detection efficiency of i-th object, $M_{V,i}$ its
luminosity, and $\delta(M_V, M_{V,i})=1$, if $M_V$ and $M_{V,i}$ are within one
bin of the histogram, and 0 otherwise.

Figure~\ref{mag_sb_vs_distance_kernel}, shows how
the maximal accessible distance depends on the galaxy luminosity (the
$r_{\rm max}(M_V)$ function).
 From this function, we can construct the
maximal accessible volume within the DR5 footprint (which covers
1/5 of the sky) as a function of galaxy luminosity, namely $V_{\rm
  max}(M_V) = 4\pi/3\,f_{DR5}\,r_{\rm max}^3(M_V)$(see
Figure~\ref{vmax}), where $f_{DR5}$ is the fraction of the sky covered
by DR5 . Then we construct the incompleteness correction $c(M_V)$,
using the probability distribution of the satellites $n(r)$. When the
maximal accessible distance for a galaxy is greater than $r_{\rm LF}$,
the correction is 1, if not it is equal to the ratio of number of satellites
within $r_{\rm max}(M_V)$ to the number of satellites within $r_{\rm LF}$:
$$c(M_V)= \left\{
\begin{array}{ll}
{\displaystyle
\frac{\int \limits_0^{r_{\rm max}(M_V)}n(r)r^2\,
dr}{\int\limits_0^{r_{\rm LF}}n(r)r^2\,dr}} & if\ r_{\rm max}(M_V)<
r_{\rm LF}  \\
1 & if\ r_{\rm max}(M_V) \geq r_{\rm LF}
\end{array}\right.$$
Finally, the luminosity function is obtained by dividing the
histogram of luminosities $h(M_V)$ by the incompleteness correction
$c(M_V)$

\end{document}